\documentclass[12pt,a4paper]{article}

\usepackage{amsmath,amsfonts,latexsym,amssymb,amscd}
\usepackage{pspicture}
\usepackage{verbatim,amsthm,curves,graphics}

\usepackage{mathrsfs}

\font\fa=bbm11

\def\ben{\begin{equation}}

\def\een{\end{equation}}

\def\bena{\begin{eqnarray}}

\def\eena{\end{eqnarray}}

\def\f(#1/#2){\frac{#1}{#2}} 

\def\Frac(#1/#2){\left(\frac{#1}{#2}\right)} 

\def\chris(#1-#2-#3){{\mit \Gamma}^{#1}{}_{{#2}{#3}} }

\def\tilchris(#1-#2-#3){\tilde{{\mit \Gamma}}^{#1}{}_{{#2}{#3}}}

\def\hatchris(#1-#2-#3){\hat{{\mit \Gamma}}^{#1}{}_{{#2}{#3}}}

\theoremstyle{definition}

\newtheorem{thm}{Theorem}

\newcommand{\myid}{{1 \!\! 1}}
\newcommand{\mr}{\mbox{\Large \fa R}}
\newcommand{\mn}{\mbox{\fa N}}
\newcommand{\mc}{\mbox{\fa C}}

\newcommand{\myint}{\mbox{$\Large \int$}}

\renewcommand{\i}{{\rm i}}
\newcommand{\supp}{\operatorname{supp}}
\renewcommand{\P}{{\mathcal P}}
\newcommand{\N}{{\mathcal N}}
\newcommand{\I}{{\mathcal I}}

\renewcommand{\L}{{\mathscr L}}
\newcommand{\F}{{\mathcal F}}
\newcommand{\T}{{\mathcal T}}
\newcommand{\D}{{\mathcal D}}

\renewcommand{\S}{{\mathcal S}}
\newcommand{\M}{{\mathcal M}}
\newcommand{\R}{{\bf R}}
\newcommand{\E}{{\bf E}}
\renewcommand{\epsilon}{\varepsilon}

\renewcommand{\O}{{\mathcal O}}
\newcommand{\WF}{\operatorname{WF}}

\renewcommand{\L}{{\mathcal L}}

\newcommand{\x}{{\vec x}}

\newcommand{\h}{{\mbox{\tiny $\rm{H}$}}}

\begin{document}

\title{The operator product expansion for perturbative
quantum field theory in curved spacetime}

\author{Stefan Hollands\thanks{\tt hollands@theorie.physik.uni-goe.de}
\\
\\
{\em Inst. f. Theor. Physik, Georg-August-Universit\"at, D-37077 G\"ottingen}
}
\maketitle

\begin{abstract}
We present an algorithm for constructing the Wilson operator product 
expansion (OPE) for perturbative interacting quantum field theory in general
Lorentzian curved spacetimes, to arbitrary orders in perturbation
theory. The remainder in this expansion is shown 
to go to zero at short distances in the sense of expectation values in 
arbitrary Hadamard states. We also establish a number of general
properties of the OPE coefficients: (a) they only depend (locally and 
covariantly) upon the spacetime metric and coupling constants, (b)
they satsify an associativity property, (c) they satisfy a
renormalization group equation, (d) they satisfy a certain microlocal 
wave front set condition, (e) they possess a ``scaling expansion''. 
The latter means that each OPE coefficient 
can be written as a sum of terms, each of which is the product of a 
curvature polynomial at a spacetime point, times a Lorentz invariant
Minkowski distribution in the tangent space of that point. The
algorithm is illustrated in an example.
\end{abstract}

\section{Introduction}

The operator product expansion~\cite{wilson} (OPE, for short) states that a 
product of $n$ local quantum fields can be expanded at short distances
as an asymptotic series, each term of which is given by a model 
dependent coefficient function
of the $n$ spacetime arguments, times a local field at a nearby reference
point\footnote{In this paper, we will take 
$y=x_n$ but other more symmetric choices are also possible.} $y$:
\ben\label{ope0}
\O_{i_1}(x_1) \O_{i_2}(x_2) \cdots \O_{i_n}(x_n) 
\sim \sum_{k} 
C_{i_1 i_2 \dots i_n}{}^k(x_1, x_2, \dots, x_n) \,  
\O_{k}(y) \, . 
\een
This
expansion has been established in perturbative quantum field theory
on Minkowski spacetime~\cite{zimmermann}, and is by now a standard
tool, for example in the analysis of quantum gauge theories such as QCD. It has
also been proven for conformally invariant quantum 
field theories~\cite{schroer,luscher,mack}, 
and has in fact played a major 
role in the developement and analysis of such theories~\cite{bpz,kac}. 
Formal mathematical 
proofs of the OPE have also been given within various axiomatic
settings~\cite{fre,bostelmann} for quantum field theory on Minkowski spacetime.
Given the importance of the OPE in flat spacetime, it is of great
interest to construct a corresponding version of the expansion 
in curved spacetime. 

In this paper, we present such a construction, within the framework of
perturbation theory. It is based 
on the perturbative construction of quantum field 
theories on globally hyperbolic Lorentzian spacetimes which 
was recently achieved in a series of papers~\cite{hw1,hw2,hw3,hw4},
which in turn were based upon key results of Brunetti, Fredenhagen,
and K\" ohler~\cite{bf,bfk}. In these papers, the interacting
(Heisenberg) quantum fields in curved spacetime are constructed as 
formal power series in the coupling constant(s), that are 
valued in a certain *-algebra of quantum observables. 
The basic idea how to construct the OPE for these interacting fields
is as follows: Suppose we have 
linear functionals $\Psi^k_y$ from the algebra into the complex
numbers, that are labelled by an index $k$ enumerating the different 
composite fields, and a reference spacetime point $y$, and which form a ``dual
basis'' to the composite quantum fields in the sense that $\Psi^k_y(\O_j(y)) =
\delta^k{}_j$. Now apply the functional with label $k$ to the OPE. 
Then we immediately see that the operator product
coefficient in front of the $k$-th term in the sum on the right side
of the OPE ought to be given precisely by the $c$-number distribution 
obtained by applying that functional to the 
product of fields on the left side of the OPE. Below, we will give a 
perturbative construction of such a dual basis of functionals in the 
context of a scalar, renormalizable field theory model in any 4
dimensional Lorentzian spacetime. In this way, we will obtain 
the desired perturbative formula for the OPE coefficients, $C$.

While this construction gives a conceptually clean derivation of 
a perturbative expression for the coefficients $C$, it does not yet
show that the remainder of the OPE expansion defined in this way
actually goes to zero (and in what sense) when the points are scaled 
together. To analyze this question, we apply any Hadamard state to 
the remainder in the OPE. We then show that the resulting $c$-number 
distribution in $n$ points goes to zero in the sense of 
distributions when the points 
are scaled together in a arbitrary fashion. Thus, the OPE holds in the
sense of an asymptotic expansion of expectation values, to arbitrary
order in perturbation theory, and for any Hadamard state. 
The proof of this statement mainly
relies on the known scaling properties of the various terms in the
perturbative series for the interacting
fields~\cite{hw1,hw2,bf}. However, these properties by themselves are
not sufficient, for the following reason. The perturbative formulae
for the interacting fields at $k$-th order involve an integration over 
$k$ ``interaction points'' in the domain of the interaction
(which might be the entire spacetime). It turns out that we can 
control the contributions from  
these integrations in the OPE if we split the interaction domain
into a region ``close'' to the points $x_1, \dots, x_n$, and a
region ``far away.'' However, the points $x_1, \dots, x_n$ themselves
are supposed to be scaled, i.e., they move, so the split of the
integration domain has to be constantly adapted. 
We achieve this by dividing the 
interaction domain into slices of thickness $2^{-j}$ centered 
about $y$, where $j=1,2,3, \dots$. We find that the 
contributions from these slices can be controlled individually, 
and then be summed, if the interaction is renormalizable. 

Finally, we will derive the following
important general properties about the OPE coefficients $C$. (a) 
They have a local and covariant dependence upon the spacetime metric,
(b) they satisfy an associativity property, (c) they satsify 
a renormalization group equation, (d) they satisfy a certain
microlocal condition on their so-called wave front set, and (e)
they can be expanded in a ``scaling expansion''. Let us explain 
these properties.

The local covariance property (a) of the OPE coefficients states that 
if $(M,g)$ and $(M',g')$ are globally hyperbolic spacetimes with 
corresponding OPE coefficients $C$ resp. $C'$, 
and if $f: M \to M'$ is a causality and orientation preserving 
isometric embedding, then $f^* C'$ is equivalent to $C$ at short
distances. Thus, in this sense, the OPE coefficients are local 
functionals of the metric (and the 
coupling constants), and in particular do not depend upon the 
large scale structure of spacetime, such as the topology of $M$.
This property can be understood from the fact that, as shown in~\cite{hw3,bf}, 
the interacting fields may be constructed in a local and 
covariant fashion. More precisely, whenever we have  
a causality preserving isometric embedding $f$,
there exists a linear map $\alpha_f$ from the quantum field algebra 
associated with $(M,g)$ into the algebra associated with $(M',g')$
preserving the algebraic relations. Furthermore, the fields are local
and covariant in the sense~\cite{hw1,bfv} 
that the image of an interacting field 
on $(M,g)$ via $\alpha_f$ corresponds precisely to the definition of 
that field on $(M',g')$. Since the OPE may be interpreted 
as an asymptotic algebraic
relation, and since the local and covariance property as just stated
means that algebraic relations only depend on the metric locally 
and transform covariantly under spacetime embedding, it is natural 
to expect that also the OPE coefficients
depend locally and covariantly upon the metric 
(and the couplings). This is indeed what we shall prove.

The associativity property (b) arises when one studies the 
different ways in which a configuration of $n$ points 
can approach the diagonal in $M^n$. For example, in the case $n=3$, 
we may consider a situation in which all three points approach each other at
the same rate, or we may alternatively consider a situation in which 
two points approach each other faster than the third one. The possible 
ways in which $n$ points can approach each other 
may be described by corresponding merger trees $\T$ which characterize
the subsequent mergers~\cite{Fultonmac, singer}. 
The associativity property states that 
scaling together the points in an operator product 
according to a given tree is equivalent
to performing subsequently the OPE in the hierarchical order
represented by the tree $\T$. We will argue---relying mainly on a 
general theorem of~\cite{hw5}---that this type of 
``short distance factorization property'' indeed holds in perturbation
theory. The associativity may again 
be understood intuitively from the fact that
the OPE coefficients are in some sense the structure ``constants''
of the abstract associative *-algebra of which the interacting fields are elements.

The renormalization property (c) of the OPE coefficients arises from the 
fact that the algebras of quantum fields satisfy a similar
property~\cite{hw3}. Namely, if we rescale the metric by a constant 
conformal factor $\lambda^2$, then this is equivalent (in the sense of
giving rise to isomorphic algebras) to redefining the field generators
in a particular way, and at the same time letting the coupling
constants of the interaction (which enter the structure of the
algebra) flow in a particular way dictated by the ``renormalization group
flow'' of the theory~\cite{hw3}. Again, since the OPE coefficients are 
in a sense the structure constants of the algebra, they can be
expected to have a corresponding property. 

The microlocal property, (d), is a property describing the nature of
the singularities in an OPE coefficient $C$. It states that the wave
front set~\cite{hormander} $\WF(C)$ has a characteristic form that encodes the
positivity of energy momentum in the tangent space. Our condition
found for $\WF(C)$ is similar in nature to the so-called 
``microlocal spectrum condition''~\cite{bfk,radzikowski} for 
the wave front set of correlation functions 
of linear field theories in curved space. However, our condition
differs qualitatively from that proposed in~\cite{bfk,radzikowski}
in that the interactions may affect the form of the wave front
set in our case.

The scaling expansion (e) states that, if we scale $n$ points $x_1, x_2, \dots, x_n$ 
in $M$ together according to a merger tree, then a given OPE
coefficient $C$ can be
approximated to any desired precision 
by a finite sum of terms each of which has the form 
of a polynomial in the mass and curvature 
tensors at point $y=x_n$, times a Lorentz invariant
Minkowski distribution in the Riemannian normal coordinates of 
$x_1, \dots, x_{n-1}$ relative to $y=x_n$. These Minkowski distributions
have, to each order in perturbation theory, a simple homogeneous
scaling behavior modified by polynomials in the logarithm. They may 
be extracted from the given OPE coefficients by taking a certain 
``Mellin moment'', which is an operation defined by first taking 
the Mellin-transform of a function and then extracting certain residue. 

The general properties just described (except (e)) are postulated
axiomatically in the fourthcoming paper~\cite{hw5}, and so our present analysis 
may be viewed as a confirmation of these axioms in perturbation theory.

\medskip

Our plan for this paper is as follows. In section 2, 
we first recall the general 
strategy for obtaining the perturbation series for the interacting 
fields~\cite{hw1,hw2,bf}. The OPE is derived in section 3, and its
general properties are derived in section 4. An example illustrating 
our algorithm for computing the OPE coefficients is presented in
section 5. 

As indicated, for simplicity and concreteness, we only consider a 
single hermitian, scalar field with renormalizable interaction (in 4
spacetime dimensions). While the restriction to a renormalizable 
interaction seems to be essential, we expect that 
our algorithm will work for other types of fields with higher spin
and renormalizable interaction, in other dimensions. However, the analysis of the OPE 
in the physically interesting case of Yang-Mills theories would first 
require an understanding of the renormalization of such theories in 
curved spacetime, which is considerably complicated by the issue of 
gauge invariance. This is at present an open problem.

\section{Perturbation theory}

A single hermitian scalar field $\phi$ in 4~dimensions is
described classically by the action 
\ben
S = \int_M \left[ g^{\mu\nu} \nabla_\mu \phi \nabla_\nu \phi +
(m^2 + \alpha R) \phi^2 + 2\sum_{i} \kappa_i \O_i \right] \, d\mu \, . 
\een
Here $d\mu = \sqrt{g} \, dx^0 \wedge \cdots \wedge dx^3$, $m^2, \alpha
\in \mr$, the quantities
$\kappa_i \in \mr$ are coupling parameters parametrizing the strength of the 
self-interaction of the field, and throughout this paper 
$\O_i$ are polynomials in the field $\phi$ and its covariant 
derivatives, as well as possibly the Riemann tensor and its
derivatives. In the above action, they encode the nature of 
the self-interaction. Later, we will assume that the interaction is
renormalizable, but for the moment no such assumption need to be made.

The perturabative construction of
the quantum field theory associated with this action has been
performed in a series of papers~\cite{hw1,hw2,hw4,bf}. These constructions
consist of the following steps. First, one defines an abstract
*-algebra~\cite{df1,hw1} $\F(M,g)$
containing the quantized field $\phi$, together with its
Wick powers $\O_i$, for the corresponding linear theory, which classically 
corresponds to dropping the self-interaction term
\ben\label{Idef}
I = \int_M \sum_{i}\kappa_i \O_i \, d\mu = \int_M {\mathcal L} \, d\mu
\, , 
\een
from the above action $S$. From these quantities, one then constructs the
corresponding interacting quantum fields (smeared with a compactly
supported testfunction) as formal power series in free
field quantities via the Bogoliubov formula~\cite{haag,kallen}, 
\ben
\O_i(h)_I = \sum_{n \ge 0} \frac{\i^n}{n!} \, \R_{n} \bigg( \myint_M
  \O_i h \, d\mu, \, I^{\otimes n} \bigg) \, , \quad h \in C^\infty_0(M)
\, , 
\een
or more formally without smearing, 
\ben\label{bog}
\O_i(x)_I = \sum_{n \ge 0} \frac{\i^n}{n!} \, \R_{n}(\O_i(x), I^{\otimes n})
\, , 
\een
where the quantities $\R_n$ are the so-called retarded products, which
are multi-linear maps on the space of local classical action
functionals of the form~\eqref{Idef},
taking values in the underlying free field algebra 
$\F(M,g)$. In order to describe these constructions 
in more detail, it is first necessary to recall at some length 
the definition and key features of the linear field algebra~$\F(M,g)$
and its quantum states, as well as nature of the retarded products $\R_n$. 

\medskip

The definition of $\F(M,g)$
can be stated in different equivalent ways~\cite{hw3,df1,df2,df3},
and we now give a definition that is most suited to our purpose. Let 
$\omega_2 \in \D'(M \times M)$ be a bidistribution of Hadamard type,
meaning (a) that $\omega_2$ is a bisolution to the Klein-Gordon equation 
$\square-m^2-\alpha R$ in each entry, (b) that the anti-symmetric part of 
$\omega_2$ is given by 
\ben\label{as}
\omega_2(x_1, x_2) - \omega_2(x_2, x_1) = \i
(\Delta_A(x_1,x_2)-\Delta_R(x_1,x_2)) \equiv \i \Delta(x_1,x_2) \, , 
\een
where $\Delta_{A/R}$ are the unique advanced and retarded propagators 
of the Klein-Gordon equation~\cite{Friedl}, and (c) that it
has a wave front set~\cite{hormander} of Hadamard type~\cite{radzikowski}:
\ben
\label{wf}
\WF(\omega_2) = \{(x_1, k_1; x_2, k_2) \in T^*(M \times M) \setminus 0;
\,\, k_1 = p, k_2 = -p, p \in \bar V_+^*\} \, .
\een
Here, it is understood that the set can only contain those $x_1$ and
$x_2$ that can be joined by a null
geodesic, $\gamma$, and that $p = p_\mu dx^\mu$ is a parallel co-vector field 
tangent to that null geodesic, meaning that $\nabla_{\dot \gamma} p =
0$, and $V^*_\pm$ is the dual of the future/past lightcone. 
The desired Wick-polynomial algebra $\F(M,g)$ is now generated by an identity 
$\myid$, and the following symbols $F$:
\ben
\label{F}
F=\int_{M^n} \left(:\phi^{\otimes n}:_\omega \, u_n \right) (x_1,
\dots, x_n) \, \bigwedge_{i=1}^n d\mu(x_i) \, , 
\een
where $u_n$ is a symmetric, compactly supported distribution on $M^n$
subject to the wave front condition~\cite{df1}
\ben\label{uclass}
\WF(u_n) \cap [(\bar V_+^*)^n \cup (\bar V_-^*)^n ] = \emptyset \, . 
\een
The relations in the *-algebra $\F(M,g)$ are as follows: The 
*-operation is defined by letting $F^*$ be given by   
the same expression as $F$, but with $u_n$ replaced by its complex conjugate,
and the product is defined by 
\begin{multline}\label{89}
:\phi^{\otimes n} :_\omega(x_1, \dots, x_n) 
:\phi^{\otimes m} :_\omega(x_{n+1}, \dots, x_{n+m}) = \\
\sum_k \frac{n!m!}{(n-k)!(m-k)!k!}
\sum_{p_1, \dots, p_k \in P} \prod_i
\omega_2(x_{p_i(1)}, x_{p_i(2)}) :\phi^{\otimes n+m-2k}:_\omega
(\{x_j; \,\, j \notin |P|\})  
\end{multline}
where $P$ is the set of all pairs $p_i \in \{1, \dots, n\}
\times \{n+1, \dots, n+m\}$. This formula is identical in 
nature to the standard Wick theorem for normal ordered quantities
(relative to a Gaussian state with 2-point function $\omega_2$). The
wave front conditions on $u_n$ and $\omega_2$ are needed in order guarantee that the product 
between the corresponding integrated quantities as in~\eqref{F}
exists, because the latter involves the pointwise products of 
distributions~\cite{hormander}. 
The definition of the algebra 
$\F(M,g)$ superficially seems to depend on the particular choice of 
$\omega_2$, but this is in fact not so: A
change of $\omega_2$ merely corresponds to a relabeling of the
generators, and does not change the definition of $\F(M,g)$ as an
abstract algebra~\cite{hw1}. 

The relation between the abstract quantization of the linear
Klein-Gordon field just described and more familiar ones is as
follows. If $f$ is a smooth test function, then the
generator~\eqref{F} with $n=1$ and $u_1=f$ 
can be identified with the 
smeared field $\phi(f)$. Indeed, using the Wick formula~\eqref{89},
and the antisymmetric part of $\omega_2$, eq.~\eqref{as}, one easily 
derives the relation $[\phi(f_1),\phi(f_2)] = \i \Delta(f_1,f_2)
\myid$, which is the standard commutation relation for a 
linear scalar field in curved spacetime. The higher order 
generators~\eqref{F} with $u_n$ given by the $n$-fold tensor product
$f^{\otimes n} = f \otimes \cdots \otimes f$ of a smooth testfunction
correspond to a smeared normal ordered product $:\phi^{\otimes
  n}:_\omega(f^{\otimes n})$, formally related to the field $\phi$ itself by  
\ben\label{nodef}
:\phi^{\otimes n}:_\omega (x_1, \dots, x_n) = \frac{\delta^n}{\i^n \delta
  f(x_1) \cdots \delta f(x_n)} {\rm exp} \left\{ \i \phi(f) +
\frac{1}{2} \omega_2(f,f)
\right\} \Bigg|_{f=0} \, .
\een
As it stands, the smeared 
version of the Klein-Gordon equation $\phi((\square - m^2 -\alpha R)f)=0$
is not an algebraic relation. However, this relation could easily be 
incorporated by factoring $\F(M,g)$ by the 2-sided *-ideal ${\mathcal
  J}(M,g)$ consisting of all elements $F$ of the form~\eqref{F} 
with $u_n$ a distribution in the class~\eqref{uclass} which is 
in the image of this distribution class under the Klein-Gordon
operator, such as $(\square - m^2 - \alpha R)f$ in the simplest case.
The purpose of this paper will be to establish an OPE, and it is 
technically convenient for this purpose not to factor by the ideal. 
However, after the OPE has been constructed there is absolutely 
no problem to factor by this ideal, because it is clear that 
the OPE will continue to hold on the factor algebra. 

Quantum states in the algebraic framework are linear
expectation functionals $\Phi: \F(M,g) \to \mc$ that are 
normalized, meaning $\Phi(\myid)=1$, and of positive type, meaning
$\Phi(F^* F) \ge 0$. Of particular importance are the so-called Hadamard
states on $\F(M,g)$. Those states are defined by the fact that 
their 2-point function $\Phi_2(x_1, x_2) = \Phi(\phi(x_1) \phi(x_2))$ 
satisfies properties (a), (b), and (c) listed above, and 
that their truncated $n$-point functions of the field $\phi$ 
for $n \neq 2$ are smooth solutions to the Klein-Gordon equation. 
The key consequence of the Hadamard requirement which
we shall need later is~\cite{hr} that $\Phi(:\phi^{\otimes n}:_\omega(x_1,
\dots, x_n))$ is smooth. Note that by definition, the $n$-point
functions of a Hadamard state satisfy the Klein-Gordon
equation. Consequently, they vanish on the ideal ${\mathcal J}(M,g)$
generated by the Klein-Gordon equation and hence induce states on the
factor algebra. 

Later, we want to define an operator product expansion, and 
for this we will need a notion of what it means for algebra elements to be
``close'' to each other. For this we now introduce a topology on
$\F(M,g)$. There are various ways to do that. A particular topology was
introduced in~\cite{hw1}. We prefer here to work with a different
(weaker) topology, defined by the collection of all linear functionals on
$\F(M,g)$ with the property that 
$\Phi(:\phi^{\otimes n}:_\omega(x_1,\dots, x_n))$ 
is smooth. This set includes the Hadamard states as defined above, and we
shall, by abuse of notation, sometimes refer to such $\Phi$ as 
``Hadamard'' as well. We then 
introduce a set of seminorms ${\mathcal N}_\Phi(F) = |\Phi(F)|$,
labelled by these functionals $\Phi$. We say that a sequence 
$\{F_N\}_{N \in \mathbb N}$ of algebra elements tends to zero if 
for each $\Phi$, and each $\epsilon>0$, there is an $N_0$
such that ${\mathcal N}_\Phi(F_N) \le \epsilon$ for all $N \ge N_0$.

An important feature of the algebra 
$\F(M,g)$ is that it has a local and covariant dependence upon 
the spacetime. More precisely, if $f: M \to M'$ is a causality 
and orientation preserving isometric embedding of a globally hyperbolic spacetime 
$(M,g)$ into another such spacetime $(M',g')$, then there exists 
a continuous, injective *-homomorphism 
\ben
\alpha_f: \F(M,g) \longrightarrow \F(M',g') \, .
\een 
This embedding is most simply described in terms of its action on a smeared
field $\phi(h)$, where $h$ is a test function on $M$. If $h'=f_* h$ is 
the corresponding pushed forward test function on $M'$, 
then we define $\alpha_f[\phi(h)] = \phi(h')$. Furthermore, the action 
of $\alpha_f$ on an arbitrary element in $\F(M,g)$ may then be
defined by continuity, because the subalgebra generated by expressions
of the form $\phi(h)$ is dense in $\F(M,g)$. The action of
$\alpha_f$ on the smeared field $\phi(h)$ is characteristic for
so-called ``local covariant fields''. Namely, an algebra valued 
distribution $\O_i: C^\infty_0(M) \to \F(M,g)$, $h \mapsto \O_i(h)$ that 
is defined for {\em all} spacetimes $(M,g)$ is called a (scalar) local and 
covariant field if 
\ben
\label{loc}
\alpha_f[\O_i(h)] = \O_i(h') \quad h' = f_* h \, , 
\een
whenever $f$ is an orientation and causality preserving 
isometric embedding. Local covariant fields of tensor type 
are defined in the same way, except that the testfunction 
$h$ is now a section in the (dual of the) vector bundle $\E_i$
corresponding to the tensor type. Thus, the field $\phi$ is 
(by definition) a local and covariant field. On the other hand, the
normal ordered $n$-th Wick power of a field defined by putting
$u_n(x_1, \dots, x_n) = f(x_n) \delta(x_1, \dots, x_n)$ in
eq.~\eqref{F} is not a local
and covariant field, because it implicitly depends on the choice of 
the 2-point function $\omega_2$, which is not a local and covariant 
quantity~\cite{hw1}. 
The possible definitions of Wick powers giving
rise to local and covariant fields (satisfying also various other
natural conditions) were classified in~\cite{hw1}. It turns out that the
definition of a given classical expression
\ben
\O_i = \nabla^{a_1} \phi \nabla^{a_2} \phi \cdots \nabla^{a_n} \phi, 
\quad i = \{a_1, a_2, \dots, a_n\} \in \I \equiv \bigoplus_n 
\mbox{\fa Z}_{\ge 0}^n 
\een
as a local, covariant field in $\F(M,g)$ is not unique, but contains certain 
ambiguities. As proven in~\cite{hw1}, 
these ambiguities correspond to the possibility of 
adding to a given field lower order Wick power powers times certain
polynomials of the Riemann tensor and its derivatives 
$\nabla_{(\mu_1} \cdots \nabla_{\mu_k)} R_{\nu_1\nu_2\nu_3\nu_4}$
of the same dimension as $\O_i$. Here, the dimension of a Wick power
is the map $[\cdot]: \I \to {\mathbb N}$ from the index set labelling
the various fields, into the natural numbers defined by 
\ben
[i] = n+\sum_{i=1}^n a_i \, .
\een
One definition which is local and covariant (and satisfies also the other 
natural conditions given in~\cite{hw1}) is the following 
``local normal ordering prescription''. It is based upon the use 
of the local Hadamard parametrix $H$, which is the bidistribution
defined on a convex normal neighborhood of the diagonal $\{(x,x); \,\, x \in M\}$
of $M \times M$ by\footnote{
The infinite sum is to be understood in the sense of an asymptotic 
expansion. 
} 
\ben
\label{Hdef}
H = \frac{v_0}{\sigma + \i 0 t} + \left( \sum_{n \ge 0} 
\frac{1}{2^n n!} 
v_{n+1} \sigma^n \right) \, \ln (\sigma + \i 0 t)\, . 
\een
In this expression, $\sigma(x,y)$ is the signed squared geodesic
distance between two points, we have defined 
$t(x,y) = \tau(x)-\tau(y)$, where $\tau$ is a time-function, 
and the $v_n(x,y)$ are smooth symmetric functions that are determined by
requiring that $H$ be a parametrix, i.e., a solution to the
Klein-Gordon equation $\square - m^2 - \alpha R$ in each entry modulo a smooth
remainder. Explicitly $v_0 = D^{1/2}/2\pi^2$ is given in terms of 
the VanVleck determinant $D$, defined by
\ben
\label{Ddef}
D = -\frac{1}{4} \frac{|(\nabla \otimes \nabla) \sigma|}{|{\mathcal J}|} \, . 
\een
Here, ${\mathcal J}(x,y) \in T_x^* M \otimes T_y^* M$
is the bitensor of parallel transport,  
$(\nabla \otimes \nabla)\sigma(x,y) \in T_x^* M \otimes T^*_y M$ is the  
is the bitensor obtained by taking the gradient of $\sigma(x,y)$
in both $x$ and $y$, and we are defining a biscalar $|{\mathcal B}(x,y)|$
by 
\ben
|{\mathcal B}(x,y)| \, d\mu(x) \otimes d\mu(y) = \bigwedge^4 {\mathcal
B}(x, y)
\een 
from any bitensor ${\mathcal B}(x,y)$.  
The smooth functions $v_n(x,y)$ are 
iteratively defined by the transport equations~\cite{deWitt} 
\ben
2(\nabla^\mu \sigma) \nabla_\mu v_n + [(\nabla^\mu \sigma)
\nabla_\mu \ln D + 4n] v_n = 
-2(\square - m^2 -\alpha R) v_{n-1} \, ,
\een
where the derivatives act on $x$. These
functions are symmetric in $x$ and 
$y$~\cite{moretti, Friedl}, and their germs at the diagonal
are locally and covariantly defined in terms of the metric. Where it
is well-defined, $H$ has a wave front set of
Hadamard type~\eqref{wf}. Next, fix a convex normal neighborhood of 
the diagonal in $M^n$, 
and in that neighborhood define locally normal ordered products
$:\phi^{\otimes n}:_\h(x_1, \dots, x_n)$ by the same formula
as~\eqref{nodef}, but with $\omega_2$ in that formula replaced by $H$, 
\ben\label{lnop}
:\phi^{\otimes n}:_\h(x_1, \dots, x_n) = \frac{\delta^n}{\i^n \delta
  f(x_1) \cdots \delta f(x_n)} {\rm exp} \left\{ \i \phi(f) +
\frac{1}{2} H(f,f)
\right\} \Bigg|_{f=0} \, .
\een
As the expressions~\eqref{nodef}, their expectation value in any 
Hadamard state is smooth. Following~\cite{hw1}
we define the local covariant $n$-th Wick power of the field as the 
distribution valued in $\F(M,g)$ given by  
\ben\label{92}
\phi^n(x) = \lim_{\epsilon \to 0} \, :\phi^{\otimes n}:_\h 
(\x(\epsilon))
\quad \x(\epsilon) = 
(\exp_x(\epsilon \xi_1),
\dots, \exp_x(\epsilon \xi_n)) \, \, ,  
\een
where $\xi_i$ denote the Riemannian normal coordinates
(identified with a vector in $\mr^4$) of the point $x_i$
relative to $x$. More generally, fields containing derivatives are defined 
by first acting with the derivatives on the appropriate
tensor factor in $:\phi^{\otimes n}:_\h$ before taking the above 
``coincidence limit''. That is, if $i=(a_1, \dots, a_n) \in \I$ denotes 
a collection of natural numbers, then the corresponding local
covariant field
$\O_i(x)$ is defined by applying the partial derivative operator
$\partial_{\xi_1}^{a_1} \cdots \partial_{\xi_n}^{a_n}$ prior to the 
coincidence limit. Note that this definition is covariant, because 
partial derivatives with respect to Riemannian normal coordinates at $x$
may be expressed in terms of curvature tensors at $x$ and covariant 
derivatives $\nabla$.

\medskip

Having described the algebra $\F(M,g)$ of field observables in the 
linear field theory associated with the action $S$ without the
interaction terms, we now turn to the interacting theory. For this 
it is technically convenient at an intermediate step to assume 
that the couplings $\kappa_i$ in the action $S$ are not constants, 
but actually smooth functions of compact support in $M$, 
which we assume are locally constant, 
\ben
\kappa_i(x) = \kappa_i \chi(x) \, ,
\een
where $\chi \in C^\infty_0(M)$, and $\chi(x) = 1$ in an open set in 
$M$ with compact closure. The cutoff functions $\chi$ serve as an infra red cutoff and are 
removed at a later stage. With the introduction of cutoff functions
understood, the interacting fields are 
defined by the Bogoliubov formula~\eqref{bog} in terms of the retarded
products $\R_n$. Each retarded product is a continuous, bilinear map
\ben
\R_{n}: \S \times \left( \bigotimes^n \S \right) \to \F(M,g)
\quad (I_0,
I_1 \otimes \cdots \otimes
I_{n}) \mapsto \R_{n}(I_0, I_1 \otimes \cdots \otimes I_{n}) \, ,
\een
from the tensor powers of the space 
$\S$ of all classical action functionals $I_j=\int {\mathcal L}_j \, d\mu$ that are 
local and polynomial in the field $\phi$ and whose couplings 
are compactly supported functions on $M$. The map is taking values in the 
algebra $\F(M,g)$ associated with the linear field theory, and is
symmetric in $I_1, \dots, I_n$.
Note that the power series expression~\eqref{bog} for the interacting fields 
is only a formal series, and no statement is made about its convergence. 
Since each term in these series is an element in $\F(M,g)$, 
the interacting fields are 
elements of the algebra $\P \otimes \F(M,g)$, where 
\ben
\P \equiv \mc[\![\kappa_1, \kappa_2, \dots ]\!] = 
\left\{ \sum_{\alpha_i \ge 0} a_{\alpha_1 \dots \alpha_k} 
\kappa^{\alpha_1}_1 \cdots \kappa_k^{\alpha_k}
;  \quad a_{\alpha_1 \dots \alpha_k} \in \mc \right\} \,  
\een 
is the corresponding ring of formal series. 
All operations, such as multiplication in this ring and in the algebra 
$\P \otimes \F(M,g)$, are 
defined by simply formally multiplying out the 
corresponding formal series term by term. Furthermore, this 
algebra inherits a natural topology from $\F(M,g)$: A formal power
series converges to another formal power series in the algebra if each 
coefficient does.

We want the zeroth order contribution $\R_0(\O_i(x))$ in the Bogoluibov
formula~\eqref{bog} to be given by our definition of 
Wick powers $\O_i(x)$ in the linear field theory\footnote{Note that
  the argument of the retarded product $\R_0(\O_i(x))$ is a {\em classical} action (or
  density), while the corresponding Wick power $\O_i(x)$ is a
  distribution valued in the quantum algebra $\F(M,g)$. We should strictly speaking
  distinguish these quantities by introducing a new notation for the Wick power, but we
shall not do this for simplicity.}. Thus, we define
$\R_0(\O_i(x))$ to be equal to the locally normal ordered 
field $\O_i(x)$ given in eq.~\eqref{92}. The terms with $n\ge 1$
in the formal series~\eqref{bog} represent the perturbative 
corrections coming from the interaction, $I$. They involve the 
higher, non-trivial, retarded products. The construction of these 
retarded products can be reduced\footnote{Alternatively, it should also be
possible to construct the retarded products directly along the lines
of~\cite{df1,df2,df3}, by suitably generalizing the arguments of 
that paper from Minkowski spacetime to curved spacetime.}
to the construction of the so-called 
``time-ordered products,'' because there exists a well-known formula
of combinatorial nature relating these two quantities, see
e.g. the appendix of~\cite{df3}. The construction of the time-ordered 
products in turn has been given in~\cite{hw1,hw2,hw4},
which is based on work of~\cite{bf}.
The strategy in these papers is 
to first write down a number of {\em functional relations} 
for the time ordered products that are motivated by 
corresponding properties of the interacting fields defined 
by the Bogoliubov formula. These properties then dictate to a large
extent the construction of the time-ordered (and hence the retarded)
products. Since there is a combinatorial 
formula relating the time ordered products to the retarded products, 
these relations can be equivalently be stated in terms of the retarded 
products. The relevant relations\footnote{A complete list
  may be found in~\cite{hw4}.} for this paper are  
as follows:

\begin{enumerate}
\item[(r1)] {\bf Causality:}
Let $F, G, S_i \in \S$. Suppose that there is a 
Cauchy surface such that $\supp G$ is 
in its future and $\supp F$ in its past. Then
\ben
\label{24}
\R_n\left( F, G \otimes \bigotimes_i S_i \right) = 0.
\een
\item[(r2)] {\bf GLZ factorization formula}~\cite{df3}:
\begin{multline}
\label{26}
\R_n \left(G, \bigotimes_i S_i \otimes F \right) - 
\R_n \left(F, \bigotimes_i S_i \otimes G \right) \\
= \sum_{I \cup J = \{1, \dots, n-1\}} \left[ \R_{|I|, 1}\left(F, 
\bigotimes_{i\in I}
S_i \right) , \, \R_{|J|, 1} \left(G, \bigotimes_{j \in J} S_j\right) \right]. 
\end{multline}
\item[(r3)] {\bf Expansion:}
There exist local covariant $c$-number distributions $r$
near the total diagonal $\{(x,x,\dots,x); \,\, x \in M\}$ 
such that (with $l_j < i_j$)
\begin{multline}
\label{rexp}
\R_n(\O_{i_0}(x), \O_{i_1}(y_1) \cdots \O_{i_n}(y_n)) = \\ 
\sum_{l_0, l_1, \dots, l_n} r^{l_0 l_1\dots l_n}_{i_0 i_1\dots
  i_n}(x,y_1,\dots,y_n) 
\, :\O_{l_0}(x) \O_{l_1}(y_1) 
\dots \O_{l_n}(y_n):_\h \, .
\end{multline}
\item[(r4)] {\bf Scaling degree:} 
The distributions $r$ have the 
scaling degree
\ben
{\rm sd}(r^{l_0 l_1\dots l_n}_{i_0 i_1\dots i_n}) = 
\sum_k [i_k] - [l_k] \, 
\een
at the total diagonal $\{(x,x,\dots,x); \,\, x \in M\} \subset M^{n+1}$. 
Here, the scaling degree of a distribution $u \in \D'(X)$
at a submanifold $Y \subset X$ is defined as
follows~\cite{steinmann,bf}. Let $S_\epsilon: X_0 \to X_0$ be 
an injective, smooth map 
defined on an open neighborhood $X_0$ of 
$Y$ with the properties (a) that $S_\epsilon \restriction_Y = id_Y$, 
and (b) that for all $y \in Y$, the map $({\rm D}S_\epsilon)(y): 
T_y X \to T_y X$ is the identity on $T_y Y$ and 
scales vectors by $\epsilon>0$ on
a complementary subspace $C_y \subset T_y X$ of $T_y Y$. Then $u$
has scaling degree ${\rm sd}(u)$ at $Y$ if $\lim_{\epsilon \to 0}
\epsilon^{D} u \circ S_\epsilon = 0$ for all $D>{\rm sd}(u)$
in the sense of $\D'(X_0)$. The definition is  independent of the 
precise choice of $S_\epsilon$. 
\item[(r5)] {\bf Locality and covariance:}
Let $f: (M,g) \to (M',g')$ be a causality and orientation
preserving isometric embedding. Then the retarded products satisfy
\ben
\label{28}
\alpha_f \left[
\R_n \left(F, \bigotimes_i S_i \right ) \right] =  
\R_{n} \left( f_* F, \bigotimes_i f_* S_i \right) \, ,  
\een
where $f_*$ denotes the natural push-forward of 
a local action functional on $M$ to the corresponding 
action functional on $M'$.
\item[(r6)] {\bf Microlocal condition:} The distributions $r$
in the expansion~\eqref{rexp} have the following wave front set:
\bena\label{wfr}
\WF(r) &\subset& \Bigg\{(x,k;y_1, l_1; \dots; y_m, l_m) \in T^*M^{m+1};
\text{there is graph in ${\mathcal G}_{1,n}$ such that}\nonumber\\
&&k=\sum_{e:s(e)=x} p_e-\sum_{e:t(e)=x} p_e, \quad l_i=\sum_{e:s(e)=y_i}
p_e-\sum_{e:t(e)=y_i} p_e \nonumber \\ 
&& y_i \in J^-(x) \quad i=1, \dots, m
\Bigg\} \, . 
\eena
The valence of the vertices $y_i$ in the graph is restricted to be 
less or equal than the maximum power of $\phi$ occurring an 
the operators $\O_{i_1}, \dots, \O_{i_n}$ in eq.~\eqref{rexp}.
\end{enumerate}
In the formulation of the last condition, we are using a graph
theoretical notation~\cite{bfk}, which will be useful later as
well. Most generally, we consider the set 
${\mathcal G}_{n,m}$ of
embedded, oriented graphs 
in the spacetime $M$ with $n+m$ vertices. 
Each such graph has $n$ so-called ``external vertices'', 
$x_1, \dots, x_n \in M$, and $m$ so-called ``internal'' or
``interaction vertices'' $y_1, \dots, y_m \in M$. These vertices are 
of arbitrary valence, and are joined by edges, $e$, which are
null-geodesic curves $\gamma_e: (0,1) \to M$. It is assumed that 
an abstract
ordering\footnote{The ordering is not assumed to be related to 
the causal structure of the manifold at this stage.} $<$ 
of the vertices is defined, and that the ordering 
among the external vertices is $x_1 < \dots < x_n$, while the 
ordering of the remaining interaction vertices is unconstrained. 
If $e$ is an edge joining two vertices, then $s(e)$ (the source) and
$t(e)$ (the target) are the two vertices $\gamma_e(0)$ and
$\gamma_e(1)$, where the curve 
is oriented in such a way that it starts at the smaller vertex 
relative to the fixed vertex ordering. Each edge carries a future
directed, tangent parallel covector field, $p_e$, 
meaning that $\nabla_{\dot \gamma_e} p_e = 0$, and $p_e \in \partial
V_+^*$.

Similar to the case of local covariant Wick products (the case
$n=0$), the above 
functional relations (together with other functional relations
described in detail in~\cite{hw1,hw2}) do not uniquely fix the
retarded products: There remains a number of real constants
at each order $n$ which parametrize the set of 
possible definitions of $\R_n$ that are compatible with (r1)--(r6). 
These correspond to the usual ``renormalization
ambiguities'' in perturbative quantum field theory,
see~\cite{hw2,hw3,hw4} for details. 

\medskip

One finally needs to remove the dependence of the interacting fields
on the arbitrary cutoff function $\chi$. For this, 
one investigates how the interacting field changes
when the cutoff function is varied. Assume that $\chi_1$ and $\chi_2$
are two different cutoff functions, both of which are equal to 1 in an open 
globally hyperbolic neighborhood $U \hookrightarrow M$. 
Let $I_1$  and $I_2$ be the corresponding interactions. Note that, as
a classical functional, 
the difference $I_1-I_2$ is supported in 
a compact region, and vanishes 
in the neighborhood $U$ where 
the cutoff functions coincide. 
The key fact~\cite{bf}, which follows from the above functional
relations (r1) and (r2), 
is now that there exists a unitary operator 
$V \in \P  \otimes \F(M,g)$, depending upon $I_1, I_2$, 
with the property that 
\ben
\label{vov}
\O_i(x)_{I_1} = V \, \O_i(x)_{I_2} \, V^*, \quad \text{for all $x \in U$, $i\in \I$.}
\een
This relation may be interpreted as saying that the algebraic relations 
between the interacting fields within the region $U$ where the cutoff function 
is constant do not depend on how the cutoff function is chosen outside this 
region, and this observation may be used to construct an abstract interacting
field algebra associated with the entire spacetime $(M,g)$
that is independent of the choice of cutoff function~\cite{bf}. However, in the present 
context, we are actually only interested in a small patch $U$ of spacetime
where we want to consider the OPE. 
Therefore, it will be more convenient
for us to simply fix an arbitrary cutoff function that is equal to 1
in the patch $U$ of interest. 
Since the OPE concerns only local algebraic relations, it is clear
from~\eqref{vov} that it should not matter what cutoff function we
choose, and this will formally be shown below in item 2) of section~4.

\section{Operator product expansion}

We will now show that the interacting fields 
$\O_i(x)_I$ described in the previous section obey an operator product 
expansion, 
\ben\label{ope}
\O_{i_1}(x_1)_I \O_{i_2}(x_2)_I \cdots \O_{i_n}(x_n)_I 
\sim \sum_{[k] \le \Delta} 
C_{i_1 i_2 \dots i_n}{}^k(x_1, x_2, \dots, x_n)_I \,  
\O_{k}(x_n)_I \, , 
\een
where $C_I$ are certain distributinal coefficients depending upon 
the interaction, $I$, which are to be determined, and where $[k]$
is the standard dimension function defined above. We mean by the above 
expression that, as the points $x_1, x_2, \dots, x_n$ approach each other, the
algebra product of the interacting fields on the left side can be
approximated, to the desired precision determined by $\Delta$, 
by the right side, in the topology on 
the algebra $\P \otimes \F(M,g)$. 

To make this statement precise, we must, however, take into account
that both sides of the OPE are actually distributional, and that a
configuration $\x = (x_1, \dots, x_n) \in M^n$ of $n$
mutually distinct points on a manifold may ``merge'' in
qualitatively different manners when $n>2$, because the points may 
approach each other at ``different rates''. The appropriate mathematical 
framework to formalize in a precise manner the possibility of 
configuration of points to approach each other at different rates 
is provided by a construction referred to as the 
``compactification of configuration space,'' 
due to Fulton and MacPherson~\cite{Fultonmac}, and Axelrod and
Singer~\cite{singer}. Let
\bena
M_0^n &=& \{\x = (x_1, x_2, \dots, x_n) \in M^n ; x_i \neq x_j \} \\
      &=& \{\x \in {\rm Map}(\{1, \dots, n\}, M), \quad \x(i) = x_i ; \quad 
            \x \, \, {\rm injective} \} \,     
\eena
be the configuration space, i.e., the space of all configurations of $n$ mutually
distinct points in $M$. The union of partial diagonals
\ben
\partial M^n_0 = \bigcup_{S \subset \{1, \dots, n\}} \Delta_S \subset M^n
\een
where a partial diagonal is defined by 
\ben
\Delta_S = \{\x \in {\rm Map}(\{1, \dots, n\}, M) ; \,\, \x \restriction S = 
{\rm constant} \}
\een
is the boundary of the configuration space $M^n_0$. 
Configurations of points where some points come close to each other 
are in some sense close to this boundary. The 
Fulton-MacPherson compactification $M[n]$ is obtained by attaching a  
different boundary, $\partial M[n]$, to $M^n_0$, which in 
addition incorporates the various directions in which $\partial M^n_0$
can be approached. 
This boundary may be characterized as the collection of endpoints of 
certain curves $\x(\epsilon)$, 
in $M[n]$, which are in $M^n_0$ for $\epsilon > 0$, and 
which end on $\partial M[n]$ at $\epsilon=0$. 
These curves are labeled by trees $\T$ that characterize 
subsequent mergings of the points in 
the configuration as $\epsilon \to 0$. A convenient way to 
describe a tree $\T$ (or more generally, the disjoint union of trees, 
a ``forest'') is by a nested set $\T = \{S_1, \dots, S_k\}$ of subsets 
$S_i \subset \{1, \dots, n\}$. ``Nested'' means that two sets are 
either disjoint, or one is a proper subset of the other. 
We agree that the sets $\{1\}, \dots, \{n\}$ are always contained in
the tree (or forest). Each set $S_i$ in $\T$ represents a node of a
tree, i.e., the set of vertices ${\rm Vert}(\T)$ is given by the 
sets in $\T$, and $S_i \subset S_j$ means 
that the node corresponding to $S_i$ can be reached by moving downward from the 
node represented by $S_j$. The root(s) of the tree(s) correspond to the 
maximal elements, i.e., the sets that are not subsets of any other
set. If the set $\{1, \dots, n\} \in \T$, then there is in fact only 
one tree, while if there are several maximal elements, then there are 
several trees in the forest, each maximal element corresponding to 
the root of the respective tree. The leaves 
correspond to the sets $\{1\}, \dots, \{n\}$, i.e., the minimal elements.  

The desired curves $\x(\epsilon)$ tending to the boundary of 
$M[n]$ are associated with trees and are constructed as 
follows. With the root(s) of the tree(s), we associate a 
point $x_i \in M$, where 
$i$ is a label that runs through the different maximal elements, 
while with each edge $e \in {\rm Edge}(\T)$ of a given tree 
(a line joining two nodes), we associate a vector 
$v_e \in T_{x_i} M$, where $x_i$ is associated with the 
root of the tree that $e$ belongs to. To describe the 
definition of this vector it is convenient to identify an edge 
$e \in {\rm Edge}(\T)$ with the pair $e=(S, S')$ of nodes 
that it connects, i.e., an edge defines a relation 
in $\T \times \T$. If $S \subset S'$, then we write
$S' = t(e)$ for target, and $S = s(e)$, for the source.
We then set
\ben\label{31}
v_e = \xi_{m(t(e))} - \xi_{m(s(e))} \quad m(S) = {\rm max}\{i; \,\, i
\in S\}\, , 
\een
where $\xi_j$ denotes the Riemannian normal coordinates
(identified with a vector in $\mr^4$ via a choice of orthonormal 
tetrad at the corresponding root $x_i$). We define the 
desired curve $\x(\epsilon)$ by
\ben
x_j(\epsilon) = 
\sum_{e \in p_j} v_e \epsilon^{{\rm depth}(t(e))}
\, .
\een
where $p_j$ is the unique path connecting the leaf $j$
with the corresponding root, 
where $v_e$ is given in terms of $\x$ by eq.~\eqref{31}, and 
where $depth(S)$ is the number of edges that connect the node $S \in
\T$ with the root. The following figure illustrates this definition
in an example:

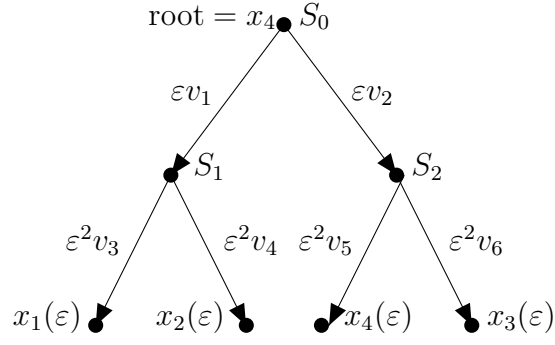
\setlength{\unitlength}{1cm}
\begin{center}
\begin{picture}(6,6.5)(0.0,0.0)
\put(3,6){\circle*{0.2}}
\put(3.2,6){$S_0$}
\put(1.2,6){${\rm root}=x_4$}
\put(1.5,4){\circle*{0.2}}
\put(1.8,4){$S_1$}
\put(4.5,4){\circle*{0.2}}
\put(4.7,4){$S_2$}
\put(0.5,2){\circle*{0.2}}
\put(-0.6,2){$x_1(\epsilon)$}
\put(2.5,2){\circle*{0.2}}
\put(1.3,2){$x_2(\epsilon)$}
\put(5.5,2){\circle*{0.2}}
\put(5.7,2){$x_3(\epsilon)$}
\put(3.5,2){\circle*{0.2}}
\put(3.8,2){$x_4(\epsilon)$}
\put(3,6){\vector(-1.5,-2){1.5}}
\put(1.5 ,5){$\epsilon v_1$}
\put(3,6){\vector(1.5,-2){1.5}}
\put(3.9 ,5){$\epsilon v_2$}
\put(1.5,4){\vector(-1,-2){1}}
\put(0.1 ,3){$\epsilon^2 v_3$}
\put(1.5,4){\vector(1,-2){1}}
\put(2.2,3){$\epsilon^2 v_4$}
\put(4.5,4){\vector(1,-2){1}}
\put(3.2 ,3){$\epsilon^2 v_5$}
\put(4.6,4){\vector(-1,-2){1}}
\put(5.2 ,3){$\epsilon^2 v_6$}
\put(1.2,1){${\mathcal T} = \{S_0, S_1, \dots, S_6\}$}
\put(-4,0.5){$x_1(\epsilon)=\epsilon v_1+\epsilon^2 v_3, \,\,
x_2(\epsilon)=\epsilon v_1 + \epsilon^2 v_4, \,\, x_3(\epsilon)= \epsilon
v_2 + \epsilon^2 v_5, \,\, x_4(\epsilon)= \epsilon v_2 + \epsilon^2 v_6$}
\end{picture}
\end{center}
For each fixed tree $\T$, and each fixed $\epsilon$, the above curve
defines a map 
\ben
\label{psidef}
\psi_\T(\epsilon): M^n_0 \to M_0^n, \quad \x \mapsto \x(\epsilon)
\een
flowing the point $\x = \x(1)$ to the point $\x(\epsilon)$. For
$\epsilon=0$, the image of this map may be viewed as a portion of the boundary 
$\partial M[n]$ corresponding to the tree. The roots 
of $\T$ correspond to the particular diagonal; in particular, 
if there is only one tree in $\T$ (as we shall assume from now on)
then the configuration $\x(\epsilon)$ converges to the total diagonal
in $M^n$. It may be
checked that the maps $\psi_\T(\epsilon)$ satisfy the composition law 
\ben
\psi_\T(\epsilon) \circ \psi_\T(\epsilon') = \psi_\T(\epsilon
\epsilon') \, .
\een
Using the maps $\psi_\T(\epsilon)$ we can define an
asymptotic equivalence relation $\sim_{\delta, \T}$ for 
distributions on $M^n$. 
Consider distributions $u_1, u_2$ defined on $M^n$. 
For a given tree $\T$ and $\delta > 0$, we 
declare the equivalence relation $\sim_{\T, \delta}$ by
\ben\label{36}
u_1 \sim_{\T, \delta} u_2 \quad :\Longleftrightarrow \quad 
\lim_{\epsilon \to 0+} \epsilon^{-\delta} \, 
(u_1 - u_2) \circ \psi_\T(\epsilon)
= 0 \,\, , 
\een
in the sense of distributions on $M^n$, where 
we view $\psi_\T(\epsilon)$ as a map $M^n \to M^n$ that is 
parametrized by $\epsilon>0$. 

Having defined the equivalence relation $\sim_{\delta, \T}$ we can now 
state precisely our notion of an OPE. Namely, we require that, for
each $\delta>0$, each given set of operators, and each tree $\T$, 
there exists a $\Delta$ so that the OPE holds in the 
sense of $\sim_{\delta, \T}$. The only issue that we have not yet been 
quite precise about is that the OPE is not a relation between 
$c$-number distributions, but instead distributions valued in 
the topological algebra $\P \otimes \F(M,g)$. This difficulty is simply dealt 
with by requiring convergence in the equivalence
relation~\eqref{36} (now for algebra valued objects) with respect to 
the topology in the algebra. Thus, we define the 
precise sense in which the OPE is supposed to hold 
to be that for each tree $\T$ with one root, and each $\delta$, there exists a
$\Delta \in \mr$ such that~\eqref{ope} holds in the sense of
$\sim_{\delta, \T}$ as a relation between the corresponding algebra valued 
distributions. 

\medskip

We now 
come to the actual construction of the operator product coefficients
in perturbation theory. 
As also described in~\cite{hw5}, and as originally suggested by
Bostelmann~\cite{bostelmann} and Fredenhagen and Hertel~\cite{fre}
in the context of algebraic quantum field theory on Minkowski spacetime, it
is convenient to think of the operator product
coefficients as arising via certain ``standard functionals''  
\ben
\Psi_{M, x}^i( \,. \,)_I: \P \otimes \F(M,g) \longrightarrow \P
\otimes \E_i |_x  \, .
\een
These functionals depend upon the given spacetime $(M,g)$, 
the  index label $i \in \I$ describing a composite field, 
a point $x \in M$, and the interaction, as indicated by 
the subscript ``$I$''. The functionals take values in the 
fiber over $x$ in the vector bundle $\E_i$ (viewed as a 
$\mathcal P$-module) associated with 
the tensor character of the field $\O_i$.
In our constructions below, 
the functionals are in fact only defined on the subalgebra 
$\P \otimes \F(U,g)$ corresponding to a convex normal neighborhood 
$U \subset M$. However, since all of our considerations
are entirely local, we may assume without loss of generality and to 
save writing that $U=M$.

The OPE coefficients are 
supposed to be given in terms of the above standard functionals by
\ben\label{Cdef}
C_{i_1 i_2 \dots i_n}{}^j(x_1, x_2, \dots, x_n)_I = 
\Psi_{x_n}^j \left( 
\O_{i_1}(x_1)_I \O_{i_2}(x_2)_I \cdots \O_{i_n}(x_n)_I 
\right)_I \, .
\een
We will construct the OPE coefficients in perturbation theory 
by presenting a suitable set of such standard functionals. 
We are going to choose 
these standard functionals as a ``dual basis'' to the interacting
fields, in the sense that we wish them to satisfy
\ben
\label{ansatz}
\Psi^i_x(\O_j(x)_I)_I = 
\delta^i{}_j \, {\rm id}_{\E_i} \quad \text{for all $x \in M$
and $[i],[j] < \Delta$.}
\een
This ansatz is motivated by the following simple consideration. 
Let us assume that an OPE exists. Let us fix a $\Delta>0$, 
carry the OPE out until $[k] \le \Delta$, and apply the functionals 
$\Psi^j_{x_n}$ to it, where $[j] \le \Delta$. Using~\eqref{ansatz}, 
we immediately find that the coefficients in the OPE must be 
given by~\eqref{Cdef}, up to a remainder term coming from the
remainder in the OPE. But this remainder is by assumption small
for asymptotically short distances, in the sense of the above 
equivalence relation, provided we make $\Delta$ sufficiently large. It
can therefore be ignored. 

Thus we have argued that if an OPE exists in the sense above, and 
if standard functionals satisfying~\eqref{ansatz} have been defined, then the 
OPE coefficients $C$ ought to be given by~\eqref{Cdef}. 
Consequently, our first step will be to 
define the standard functionals as formal power series in the 
coupling constants $\kappa_i$ so that eq.~\eqref{ansatz} will be
satisfied to arbitrary orders in perturbation theory. To zeroth order 
in perturbation theory, such standard functionals are defined 
as follows (see also~\cite{hw5}). Recall that a general 
algebra element $F \in \F(M,g)$ can be written as in eq.~\eqref{F}
in terms of normal ordered generators~\eqref{nodef}. If we are
interested only in elements $F$ so that the corresponding $u_n$
in~\eqref{F} are suppored sufficiently close to the diagonal in $M^n$
(as we will always assume in the following), then we may rewrite $F$
in terms of the locally normal ordered generators $:\phi^{\otimes
  n}:_\h(x_1, \dots, x_n)$ given in eq.~\eqref{lnop}
instead of the normal ordered generators
$:\phi^{\otimes n}:_\omega(x_1, \dots, x_n)$. The action of 
the zeroth order standard functionals is then declared by
\ben\label{standard0}
\Psi^i_x\left( :\phi^{\otimes m}:_\h(x_1, \dots, x_m) \right)
= \frac{\delta_{m,n}}{a_1! \cdots a_n!} \xi_1^{\otimes a_1}
\cdots \xi^{\otimes a_n}_n \quad i = (a_1, \dots, a_n) \, , 
\een
and extended to all of $\F(M,g)$ by linearity.
Here, $\xi_i$ are the Riemannian normal coordinates of $x_i$ relative
to $x$, identified with vectors in $T_x M$. 
These functionals satisfy the analog of
eq.~\eqref{ansatz} for the linear fields defined above. Since the
interacting fields $\O_i(x)_I$ are given by formal power series 
whose zeroth order is the linear field expression (see the Bogoliubov 
formula~\eqref{bog}), it follows that the action of the linear field 
functionals on an interacting field is of the form
\ben
\Psi_x^i(\O_j(x)_I) = \delta^i{}_j \, {\rm id}_{\E_i} + A^i{}_j(x) \, , 
\een
where $A^i{}_j$ is the endomorphism in ${\rm End}(\E_j, \E_i)$
that arises from the higher perturbative
contributions to the interacting field, see~\eqref{bog}, and is given 
by 
\ben
A^i{}_j(x) = \sum_{n \ge 1} \frac{\i^n}{n!} \Psi^i_x \left(
\R_n(\O_j(x); I^{\otimes n})
\right) \, .
\een 
Consequently, using the 
standard geometric series for the inverse of a linear operator of the form 
$\myid + L$ and writing out explicitly the above formula for 
$A^i{}_j(x)$, we find that the functional 
defined by the following series is a solution to the 
equation~\eqref{ansatz}: 
\begin{multline}
\Psi^i\left( F \right)_I = \sum_{k=0}^\infty (-1)^k 
\sum_{m_l \ge 1} \frac{\i^{m_1 + \dots + m_k}}{m_1! \cdots m_k!} \\
\Psi^i\left(\R_{m_1}(\O_{j_1}; I^{\otimes m_1}) \right)
\Psi^{j_1} \left(\R_{m_2}(\O_{j_2}; I^{\otimes m_2}) \right)
\cdots
\Psi^{j_k}(\R_{m_k}\left(\O_{j_{k+1}}; I^{\otimes m_k}) \right)
\,\,  
\Psi^{j_{k+1}}\Big( F \Big) \, .  
\end{multline}
Here, $m=\sum m_l$ is the perturbation order of an individual
term, and the sums over $j_l$ are carried out to order $[j_l] \le
\Delta$. Thus, for each fixed $m$, the sum over $k$ has only a finite 
number of terms, and the resulting expression is a well-defined 
functional on formal power series, valued in formal power series. 
Furthermore, all functionals $\Psi^{j_k}$ and all operators $\O_{j_k}$ appearing on 
the right side are taken at a reference point $x$. We now define the 
operator product coefficients by formula~\eqref{Cdef} in terms 
of the functionals  $\Psi^i( \, . \,)_I$. Writing out all terms
explicitly, the interacting OPE-coefficients are thus given by  
\begin{multline}\label{cijk}
C_{i_1 \dots i_n}{}^j(x_1, \dots, x_n)_I \equiv \sum_{k=0}^\infty (-1)^k 
\sum_{m_i \ge 1} \frac{\i^{m_1 + \dots + m_k}}{m_1! \cdots m_k!}
\sum_{[l_j]\le \Delta}\\
\Psi^j \left(\R_{m_1}(\O_{l_1}(x_n); I^{\otimes m_1}) \right)
\Psi^{l_1} \left( \R_{m_2}(\O_{l_2}(x_n); I^{\otimes m_2}) \right)
\cdots
\Psi^{l_k}(\R_{m_k} \left( \O_{l_{k+1}}(x_n); I^{\otimes m_k}) \right) \\
\times \sum_{n_i \ge 0} \Psi^{l_{k+1}} 
\left( \prod_{r=1}^n \frac{\i^{n_r}}{n_r!}\R_{n_i}(\O_{i_r}(x_i); I^{\otimes n_r}) \right) \, ,   
\end{multline}
where all local functionals $\Psi^l$ refer to the point $x_n$.
In order to make this formula well-defined, it is necessary to 
assume that the support of the cutoff function $\chi$ implicit in 
$I$ is small enough so that the standard functionals are defined on
the corresponding retarded product. However, this is no real
restriction, because the OPE is an asymptotic short distance expansion, and 
we will later show that the coefficients do not depend on the
particular choice of $\chi$ asymptotically.

We claim that the coefficients $C_I$ satisfy an OPE:
\begin{thm}
Let the interaction $I=\int {\mathcal L} \, d\mu$ be renormalizable,
i.e., $[{\mathcal L}] \le 4$.
For a given tree $\T$, $\delta \ge 0$, and given $i_1, \dots, i_n \in \I$, let 
\ben
\Delta = \delta + \left( \sum_{j=1}^n [i_j] \right) 
\cdot {\rm depth}(\T)\, , 
\een
and define the OPE coefficients $C_{i_1 \dots i_n}{}^k$ by
eq.~\eqref{cijk}. Then the OPE holds:
\ben\label{ope1}
\O_{i_1}(x_1)_I \O_{i_2}(x_2)_I \cdots \O_{i_n}(x_n)_I 
\sim_{\T,\delta} \sum_{[k] \le \Delta} 
C_{i_1 i_2 \dots i_n}{}^k(x_1, x_2, \dots, x_n)_I \,  
\O_{k}(x_n)_I \, . 
\een
\end{thm}
\paragraph{\bf Remarks:} 1) The theorem is false for
non-renormalizable interactions. \\
2) Since the topology on $ \P \otimes
\F(M,g)$ of which the interacting fields are elements is generated by 
a set of seminorms associated with functionals including the Hadamard
states, it follows that the OPE will continue to hold on the factor 
algebra obtained by dividing by the Klein-Gordon equation, in the 
sense of expectation values in Hadamard states, to arbitrary orders in
perturbation theory.

\medskip
\noindent
{\it Proof:}
Let $\N_I$ be defined as the remainder in the OPE, i.e., the left side
of~\eqref{ope1} minus the right side.  We need to prove 
that $\epsilon^{-\delta} \Phi(\N_I \circ \psi_\T(\epsilon))$ tends to
$0$ in the sense of distributions as $\epsilon \to 0$. 
The analysis of this limit is easiest in the case when $\T$ 
is the tree $\T=\{S_0, S_1, \dots, S_n\}$ with one root 
$S_0=\{1, \dots, n\}$ and 
$n$ leaves $S_i = \{i\}$. Then $depth(\T)=1$, and $\psi_\T(\epsilon)$
is the map that scales the Riemannian normal coordinates of the 
points $x_1, \dots, x_{n-1} \in U$ relative to $y = x_n$ by
$\epsilon$, where $U$ is a convex normal neighborhood of $y$. 
Thus, taking $\epsilon = 2^{-N}$, we must show that 
\ben\label{2N}
2^{\delta N} \Phi(\N_I(2^{-N} x_1, \dots, 2^{-N} x_n)) 
\quad \text{as $N \to \infty$}, 
\een
in the sense of distributions valued in $\mathcal P$, i.e., to any 
order in perturbation theory. In the above expression, 
and in the remainder of this proof, points $x_i$ have been identified
with their Riemannian normal coordinates around $y = x_n$. In
order to analyze the above expression, it is necessary to perform several 
intermediate decompositions of $\N_I$, and we now explain how this is
done. 

We first decompose $\N_I$ into contributions 
from the different orders in perturbation
theory. The ring $\mathcal P$ of formal power series in the couplings 
$\kappa_i$ contained in the interaction $I = \int \L d\mu$ can be
decomposed into a direct sum 
\ben
{\mathcal P} = \bigoplus_k {\mathcal P}_{(k)}, \quad 
{\mathcal P}_{(k)} = \text{Eigenspace of $\sum \kappa_i \, d/d
  \kappa_i$ for eigenvalue $k$},  
\een
where the $k$-th summand corresponds to the $k$-th order in
perturbation theory. Accordingly, $\N_I$ may be decomposed as 
$\sum \N_{(k)}$ into contributions from the various orders in
perturbation theory, and likewise $C_I = \sum C_{(k)}$ etc. Using  
the Bogoliubov formula, the $k$-th
order perturbative contribution to $\N_I$ can be written in the form 
\bena
\N_{(k)}(x_1, \dots, x_n) 
&=& \sum_{k_1+\dots+k_n=k} \frac{\i^k}{k_1! \cdots k_n!}
\prod_j \R_{k_j}(\O(x_j), I^{\otimes k_j}) \nonumber\\
&-& \sum_{p=0}^k C_{(p)}(x_1, \dots, x_n) \R_{k-p}(\O(x_n), I^{\otimes
  (k-p)}) \, , 
\eena
where the labels on $\O$ indicating the field species have been 
omitted to lighten the notation.
Let $A_{(k)}$ be defined as $\N_{(k)}$, but with the $k$-th order 
OPE coefficient $C_{(k)}$ omitted. Then, using the definition of the 
OPE-coefficients, it can be seen that 
\ben
C_{i_1 \dots i_n}{}^j(x_1, \dots, x_n)_{(k)} = \Psi^j(
A_{i_1 \dots i_n}(x_1, \dots, x_n)_{(k)})
\een
and that $\N_{(k)} = A_{(k)} - \sum_{[j] \le \Delta} \Psi^j(A_{(k)})
\O_j$, where $\Psi^j$ are the free field reference functionals 
at point $x_n$ and where $\O_j$ are the free field Wick powers
taken at point $x_n$. Thus, the expectation value 
of the scaled, $k$-th order perturbative contribution to the remainder 
is given by 
\begin{multline}\label{NA}
\Phi(\N_{(k)}(2^{-N} 
x_1, \dots, 2^{-N} x_n)) = 
\Phi(A_{(k)}(2^{-N} x_1, \dots, 2^{-N} x_n))\\ - 
\sum_{[j] \le \Delta} \Psi^j (A_{(k)}(2^{-N} x_1, \dots, 2^{-N} x_n))
\Phi(\O_j(2^{-N} x_n)) \, , 
\end{multline}
The right side of this equation is schematically of the form 
$\Phi(F) - \sum_k \Psi^k_y(F) \Phi(\O_k(y))$, and for such 
expressions we will now write down an expression which will be useful 
to analyze the limit $N \to \infty$ of eq.~\eqref{NA}. 
To derive this expression, 
perform a Taylor expansion with remainder about $(y,\dots,y) \in U^m$ of the 
$m$-th locally normal ordered
product~\eqref{lnop},  
\begin{multline}
: \prod_{i=1}^m \phi(\xi_i) :_\h - 
\sum_{|\alpha_1|+\dots+|\alpha_m|\le\rho} \frac{1}{\alpha_1!
  \cdots \alpha_m!} : \prod_{i=1}^m \xi_i^{\alpha_i}
\partial^{\alpha_i} \phi(0) :_\h \\
=\frac{1}{\rho!} \int_0^1 (1-t)^{\rho} \, \partial_t^{\rho+1} \,
:\prod_{i=1}^m \phi(t\xi_i) :_\h \, dt \, .
\end{multline}
Here, the $\xi_i$ denote Riemannian normal coordinates around $y$ and are 
identified with points in $\mr^4$,  
$\alpha_i \in {\mn}_0^4$ is a multiindex, and quantities like
$|\alpha_i|$ are defined using standard multiindex conventions. 
As explained above, any element $F \in \F(M,g)$ supported in $U$ 
may be written as a linear
combination of expressions which consist of distributions $u_m$ 
supported in $U^m$ satisfying the wave front condition~\eqref{uclass}, 
integrated with locally normal ordered products $:\phi^{\otimes
  m}:_\h$. If we apply a Hadamard state $\Phi$ to such an expression $F$, 
use the above Taylor series with remainder, 
and use the definition~\eqref{standard0} for the standard
functionals, then we get the following equation:
\begin{multline}\label{95}
\Phi\left( F \right) - \sum_{[k]<\Delta} \Psi_{y}^k(F) 
\, \Phi(\O_k(y))\\
= \sum_m \sum_{|\alpha| = \Delta-m+1}\, 
\frac{1}{(\Delta-m)!} \int_{M^m} 
u_m(\xi_1, \dots, \xi_m) \xi_1^{\alpha_1} \cdots \xi^{\alpha_m}_m
\\ \int_0^1 (1-t)^{\Delta-m} \Phi \left(
: \partial_{\alpha_1} \phi(t\xi_1) \cdots \partial_{\alpha_m} (t\xi_m)
:_\h 
\right) dt
\, \bigwedge_{i=1}^m d\mu(\xi_i)  \, .  
\end{multline}
The key point to note about this identity is that 
there are now factors of $\xi_i^{\alpha_i}$ on the right side, 
which will work in our advantage when the points $\xi_i$ are scaled 
by a small factor. On the other hand, the normal ordered expectation 
values in the second line are smooth (here we are using the assumption
that $\Phi$ is Hadamard), and so will not cause any trouble for such a scaling.
We will now prove that~\eqref{2N} holds in the sense of 
distributions by exploiting this
identity for $F=A_{(k)}$ in eq.~\eqref{NA}. 
However, before we efficiently make use of that identity in~\eqref{NA}, it 
is first necessary to rewrite $A_{(k)}$ in a suitable way, and to
apply an induction in $k$. 

For this, we recall that the interaction 
Lagrangian density $\L$ is confined to the convex normal neighborhood 
$U$ since we are taking the couplings to be 
$\kappa_i(x) = \kappa_i \chi(x)$ with $\chi$ a smooth cutoff function 
that is supported in $U$. We now ``slice up'' the support 
of $\L$ into contributions
from different ``shells'' in $U$ that are centered around $y = x_n$, 
and that have thickness $2^{-j}$, where $j=1, \dots, N$. For this, 
we choose a compactly supported function $\vartheta$ that is 1 on 
$U$, and we set
\ben
\vartheta_j(x) = \vartheta(2^j x) \, .
\een
Then $\L$ may be decomposed as 
\ben
\L = \vartheta_N \L + \sum_{j=1}^N (\vartheta_{j-1}-\vartheta_{j}) \L \,
. 
\een
Each term in the sum is supported in a slice of thickness
$2^{-j}$, see the following figure on p.~24. 
The key step is now to rewrite an interacting field quantity
in a way that reflects the subdivision of the interaction region $U$
into these slices. For this, we note that if $V_j$ is the unitary 
in~\eqref{vov} relating the interacting field with interactions
$I_j = \int \vartheta_j \L d\mu$ and $I_{j-1} = \int \vartheta_{j-1}
\L d\mu$, we have 
\ben\label{shell}
\O_I(2^{-N}x) = V_1 V_2 \cdots V_N \O_{I_N}(2^{-N} x) (V_1 V_{2}
\cdots V_N)^{-1} \, ,  
\een
for all $x \in U$. Explicitly, $V_j$ is given in terms of the relative 
S-matrix~\cite{bf}, 
\ben
V_j = S_{\myint \vartheta_j \L}\left( 
\myint \rho_j \L
\right) 
= \sum_k S_{\myint \vartheta_j \L}\left( 
\myint \rho_j \L
\right)_{(k)} \, .  
\een
Here, $\rho$ is any smooth function of compact support in $U$ with the
property that $\rho(x) = 0$ for all $x \in J^+(\supp(\vartheta_1))$ and 
$\rho(x) = \vartheta_0(x) - \vartheta_1(x)$ for all $x \in
J^-(\supp(\vartheta_1))$, and $\rho_j(x) = \rho(2^j x)$. 
Each term in this expansion can in turn be written in terms 
of retarded products~\cite{bf}.
Substituting the equation~\eqref{shell} into the formula for the
remainder, and expanding in a perturbation expansion, we get the
following identity:
\begin{multline}\label{Ak}
A_{(k)}(2^{-N} x_1, \dots, 2^{-N} x_n)_{I} = 
A_{(k)}(2^{-N} x_1, \dots, 2^{-N} x_n)_{I_N} \\ 
+ \sum_{p=0}^{k-1} \sum_{
\scriptsize{
\begin{matrix}
k-p\!\!\!\! \!\!\! &=k_1+ \dots +k_r\\
                   &+l_1+ \dots +l_q 
\end{matrix}}
} 
\prod_{0 < \alpha_1 \cdots < \alpha_r < N} 
S_{\myint \vartheta_{\alpha_j} \L}\left( 
\myint \rho_{\alpha_j} \L
\right)_{(k_j)} \\
\cdot \N_{(p)}(2^{-N} x_1, \dots, 2^{-N} x_n)_{I_N} \\
\cdot \prod_{0 > \beta_1 \cdots > \beta_q > N} 
S_{\myint \vartheta_{\beta_i}\L}\left( 
\myint \rho_{\beta_i} \L
\right)_{(l_i)}^* \, .
\end{multline}
This complicated identity has the following structure. The sum 
on the right side is by definition 
only for $p$ such that $p < k$, meaning
that the terms under the sum only contain the remainder in the OPE 
up to $(k-1)$-th order in perturbation theory. This will enable us 
to use an inductive procedure to estimate the $k$-th order
perturbative contribution to the remainder by the lower order
contribution. The first term on the right side,
$A_{(k)}(\dots)_{I_N}$, is identical in nature with
$A_{(k)}(\dots)_{I}$, with only exception that all the retarded
products implicit in its definition are now computed with respect to 
the interaction $I_N = \int \vartheta_N \L d\mu$ which is supported
only in a small ball of radius $2^{-N}$ around $y=x_n$. This will
enable us to use a scaling argument to estimate this term.  

We now explain more precisely 
how the decomposition of $A_{(k)}$ given in~\eqref{Ak}
will make it possible to analyze the scaling behavior of the $k$-th 
order remainder in the OPE. For this, we take~\eqref{Ak}, 
and substitute it into eq.~\eqref{NA}. This gives an expression for 
$\Phi(\N_{(k)}(\dots)_I)$, and to each term in this expression, we can apply
eq.~\eqref{95}. Consider first the term arising from the first term 
on the right side of eq.~\eqref{Ak}. That term makes a contribution 
to $\Phi(\N_{(k)}(\dots)_I)$ of the form 
\ben\label{Akk}
\Phi(A_{(k)}(2^{-N} x_1, \dots, 2^{-N} x_n)_{I_N}) - 
\sum_{[j] \le \Delta} \Psi^j (A_{(k)}(2^{-N} x_1, \dots, 2^{-N} x_n)_{I_N})
\Phi(\O_j(2^{-N} x_n)_{I_N}) \, .  
\een
We must now substitute the expression for $A_{(k)}(\dots)_{I_N}$. 
For this, we use that, on account of the Bogoliubov formula, 
the scaled interacting field with interaction $I_N$ is given by 
\ben
\O(2^{-N} x)_{I_N} = \sum_k \frac{2^{-4kN}\i^k}{k!} \int_{U^k} 
\R_k \left(\O(2^{-N}x), \bigotimes_{i=1}^k 
\L(2^{-N} y_i) \right)
\prod \vartheta_0(y_k) d \mu_N(y_k) \, , 
\een
where we have performed the change of integration variables
$y_i \to 2^{N} y_i$, and where $d\mu_N$ is $2^{4N}$-times the 
pull-back of $d\mu$ by the inverse of this map, which is smooth as $N\to \infty$. 
If we now also use the Wick expansion of 
the retarded products (r3), along with the 
scaling degree property~(r4) and combine the result with 
eq.~\eqref{95}, then we obtain that eq.~\eqref{Akk} scales as
$2^{-N(\delta+1)}$, as desired. 

Now we must take the second term on the right side in~\eqref{Ak},
substitute it into~\eqref{NA}, and then analyze its scaling
using~\eqref{95}. To do this, we must now proceed iteratively, in 
the order in perturbation theory $k$. For $k=0$, there is nothing to
show. For $k>0$, we then inductively know the
scaling~\eqref{2N} up to order $k-1$, which enables us to estimate the 
remainders $\N_{(p)}, p<k$ in the terms in the sum on the right side
of~\eqref{Ak}. More precisely, we may inductively assume that the
$p$-th order remainder ($p<k$) has the structure
\begin{multline}\label{Np}
\N_{(p)}(2^{-N} x_1, \dots, 2^{-N} x_n)_{I_N} =  
\sum_m 2^{-N(4m+\Delta-m+1)}\sum_{|\alpha| = \Delta-m+1}\, 
\frac{1}{(\Delta-m)!} \\
\cdot \int_{M^m} 
n_m(2^{-N} y_1, \dots, 2^{-N} y_m, 2^{-N} x_1, \dots, 2^{-N} x_n) 
y_1^{\alpha_1} \cdots y^{\alpha_m}_m
\\ \int_0^1 (1-t)^{\Delta-m} \
: \partial_{\alpha_1} \phi(t2^{-N}y_1) \cdots \partial_{\alpha_m} (t2^{-N}y_m)
:_\h dt
\, \bigwedge_{i=1}^m d\mu_N(y_i)  \, , 
\end{multline}
where $n_m$ are the coefficients in a Wick-expansion of
$\N_{(p)}(x_1, \dots, x_n)_{I}$ (note that we have also performed a change of 
integration variables $y_i \to 2^N y_i$). 
Using the scaling (r3) and the fact that all
terms $\N_{(p)}(\dots)_I$ may be written in terms of retarded products 
by means of the Bogoliubov formula, one can see that 
\ben\label{nscale}
2^{-N([j_1]+\dots +[j_n] + m[\L] -m)} 
n_m(2^{-N} y_1, \dots, 2^{-N} y_m, 2^{-N} x_1, \dots, 2^{-N} x_n) 
\to 0 \quad \text{as $N \to \infty$.}
\een
One now has to take formula~\eqref{Np}, and substitute it into the sum 
on the right side of~\eqref{Ak}. From the product of
$\N_{(p)}(\dots)_{I_N}$ with the relative $S$-matrices there arise 
terms which blow up as $N \to \infty$, and so these terms have to be 
carefully controlled. To understand in detail what type of diverging
terms can arise, we must write out the explicit formula for the 
relative $S$-matrices in terms of retarded products. Then we must write 
each retarded product in a Wick expansion (r4), and perform the 
products using Wick's theorem~\eqref{89}, with $\omega$ replaced by $H$. 
Then we get a collection of terms, each of 
which is a product of $H$, $r$, $n_m$ and a locally normal ordered 
Wick power. These terms are evaluated on a set of
spacetime arguments which are scaled by $2^{-N}, 2^{-\alpha_i}$, or
$2^{-\beta_j}$, and which are integrated against the compactly
supported smooth functions $\vartheta$ or
$\rho$. The arguments 
scaled by $2^{-\alpha_j}$ arise from points in the interaction domain
$U$ within the $\alpha_j$-th slice, the arguments 
scaled by $2^{-\beta_j}$ arise from points in the interaction domain
$U$ within the $\beta_j$-th slice, and the arguments scaled by $2^N$
correspond to the scaled arguments $2^{-N} x_i$. More precisely, when we 
use Wick's theorem to perform the products in the second
term in~\eqref{Ak}, there arise ``contractions'' between
points in the $\alpha_i$-th and $\beta_j$-th slice, indicated
by lines in the following figure:

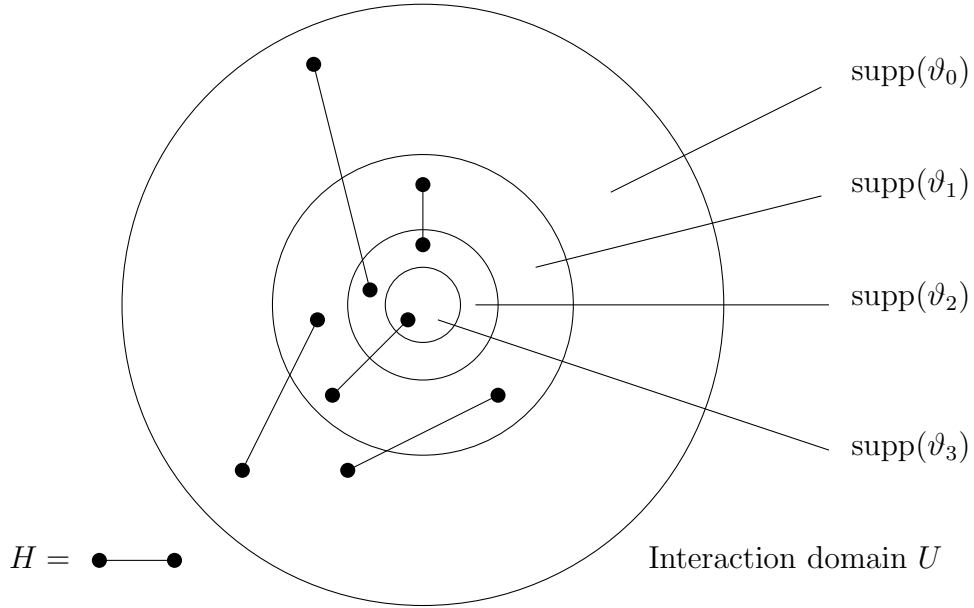
\setlength{\unitlength}{1cm}
\begin{center}
\begin{picture}(6,8)(0.0,0.0)
\put(3,4){\circle{1}}
\put(3,4){\circle{2}}
\put(3,4){\circle{4}}
\put(3,4){\circle{8}}
\put(3.2,3.8){\line(3,-1){5.2}}
\put(3.7,4){\line(1,0){4.7}}
\put(4.5,4.5){\line(4,1){3.8}}
\put(5.5,5.5){\line(2,1){2.8}}
\put(8.7,4){$\supp(\vartheta_2)$}
\put(8.7,5.5){$\supp(\vartheta_1)$}
\put(8.7,7){$\supp(\vartheta_0)$}
\put(8.7,2){$\supp(\vartheta_3)$}
\put(6,0.5){Interaction domain $U$}
\put(-2.5,0.5){$H = $}
\put(-1.3,0.6){\line(1,0){1}}
\put(-1.3,0.6){\circle*{0.2}}
\put(-0.3,0.6){\circle*{0.2}}
\put(2.8,3.8){\circle*{0.2}}
\put(1.8,2.8){\circle*{0.2}}
\put(2.8,3.8){\line(-1,-1){1}}
\put(3,4.8){\circle*{0.2}}
\put(3,5.6){\circle*{0.2}}
\put(3,4.8){\line(0,1){0.8}}
\put(4,2.8){\circle*{0.2}}
\put(2,1.8){\circle*{0.2}}
\put(4,2.8){\line(-2,-1){2}}
\put(1.6,3.8){\circle*{0.2}}
\put(0.6,1.8){\circle*{0.2}}
\put(1.6,3.8){\line(-1,-2){1}}
\put(2.3,4.2){\circle*{0.2}}
\put(1.55,7.2){\circle*{0.2}}
\put(2.3,4.2){\line(-1,4){0.75}}
\end{picture}
\end{center}

Each such contraction is associated with a factor $H(2^{-\alpha_j} y_1,
2^{-\beta_i} y_2)$ (or a derivative thereof),
which, using the explicit form of the Hadamard parametrix $H$, is seen
to scale as $2^{2 min(\alpha_i,\alpha_j)}$, in the sense of the 
scaling degree of a distribution (with a correspondingly larger power
when derivatives are present). Furthermore, 
the scaling of the retarded products 
in a term in eq.~\eqref{Ak} associated with the $i$-th slice may also 
be controlled. Namely, using the Wick expansion (r3), we see that a 
retarded product associated with the $i$-th slice contributes factors
of $r(2^{-i} y_1, \dots, 2^{-i} y_l)$, the scaling power of which 
may then be controlled using (r4). Finally, the scaling of 
$n_m(2^{-N} x_1, \dots, 2^{-N} y_m)$ is controlled by eq.~\eqref{nscale}.
Thus, the rate at which the terms in the sum on the right side of 
\eqref{Ak} blow up can be controlled. We finally need to substitute each 
such term into eq.~\eqref{NA}, and use~\eqref{95}. 
If we carefully keep track of all the 
scaling powers, then we find that a typical 
term contributing to $\N_{(k)}$ arising from these substitutions
has the scaling power 
\ben
2^{N(-\delta-4k-1)+ [\L] \sum_{j} k_j \alpha_j + [\L]\sum_i l_i \beta_i}
\een 
Using a geometric series, the sum of such terms is estimated by 
\begin{multline}
\sum_{p=0}^{k-1} \sum_{
\scriptsize{
\begin{matrix}
k-p\!\!\!\!\!\!\! 
&=k_1+ \dots +k_r\\
       &+l_1+ \dots +l_q 
\end{matrix}}
}
\sum_{
\scriptsize{
\begin{matrix}
 1<\alpha_1 \dots <\alpha_r<N \\
 1<\beta_1 \dots <\beta_q<N
\end{matrix}}
}
2^{N(-\delta-4k-1) + [\L] \sum_{j} k_j
  \alpha_j + [\L]\sum_i l_i \beta_i} \\
\le {\rm const.}\, 2^{-N(\delta+1-([\L]-4)k)} \le {\rm const.}\, 2^{-N(\delta+1)}, 
\end{multline}
where we have used in the last step that the interaction is
renormalizable, $[\L] \le 4$. Thus, the total scaling of the sum of 
terms in $\N_{(k)}(2^{-N} x_1, \dots, 2^{-N}x_n)$ is given by
$2^{-N(\delta+1)}$, which implies the convergence of~\eqref{2N}. On the
other hand, for non-renormalizable interactions, we would not get 
convergence.

The analysis for a general tree $\T$ is in principle not very
different from the one just given. For a general tree, the minimum
distance between points in a scaled configuration $\x(2^{-N})$ is of 
order $2^{-depth(\T) N}$, and not just $2^{-N}$ as in the simple tree
studied above. This implies that the scaling of the corresponding
quantities in the perturbative expansions is different. 
One now has to go through the above steps
again and take that different scaling into account. If this is done, 
then the result claimed in the theorem is obtained. \qed

\section{Properties of the OPE coefficients}

We would now like to establish a number of important general
properties of the OPE coefficients defined in the previous
section. These properties are
\begin{enumerate}
\item Microlocal spectrum condition.
\item Locality and covariance.
\item Renormalization group.
\item Associativity.
\item Scaling expansion.
\end{enumerate}
Except for the last one, these properties were postulated as axioms in the paper~\cite{hw5}, 
so our present work can be viewed as a confirmation of~\cite{hw5}. 

We first establish the microlocal spectrum condition. With the graph 
theoretical notation ${\mathcal G}_{n,m}$ introduced in section~2, let 
us define the following subset of the cotangent space $T^*M^n$:
\begin{eqnarray}
\label{gamtdef}
\Gamma_{n,m}(M, g) &=& 
\Bigg\{(x_1, k_1; \dots; x_n, k_n) \in T^*M^n \setminus \{0\} ; \,\, 
\exists \,\, \text{decorated graph $G(\x, \vec y, \vec p) \in
  {\mathcal G}_{m,n}$} \nonumber\\
\vspace{1cm}
&& \text{such that
$k_i = \sum_{e: s(e) = x_i} p_e - \sum_{e: t(e) = x_i} p_e$ for all
$x_i$ and}\nonumber\\
&& \text{such that
$0 = \sum_{e: s(e) = y_i} p_e - \sum_{e: t(e) = y_i} p_e$ for all
$y_i$ and} \nonumber\\
&& \text{such that $y_i \in J^+(\{x_1, \dots, x_n\}) \cap J^-(\{x_1,
  \dots, x_n\})$ for all $1 \le i \le m$,} 
\Bigg\} \, . \nonumber\\ 
\end{eqnarray}
The microlocal spectrum condition for the OPE-coefficients is statement that
\ben
\label{msc}
\WF(C) \restriction U_n \subset \overline{\bigcup_{m \ge 0} \Gamma_{m,n}} \, . 
\een
where $U_n$ is some neighborhood of $\{(x,x,\dots,x) \in M^n\}$, 
and where ``WF'' is the wave front set of a distribution~\cite{hormander}.

The microlocal condition in the above form~\eqref{msc} is similar in 
nature to a condition that was obtained by~\cite{bfk} for the $n$-point
correlation functions of Wick powers in Hadamard states in the context 
of linear field theory. The difference to the above condition is that 
also interaction vertices are now allowed, which were not considered
in~\cite{bfk}. These interaction vertices correspond to the
contributions $m\ge 1$ in~\eqref{msc} and genuinely weaken the bound
on the wave front set relative to the linear case (for $n \ge 4$). The 
interaction vertices arise from the non-linear 
interactions present in the theory. As we will see, 
the maximum valence of the interaction 
vertices allowed in $\WF(C)$ is equal to the maximum power of the 
field $\phi$ that appears in the interaction Lagrangian $\mathcal L$. 
Since we restrict ourselves to renormalizable interactions in 4
spacetime dimensions, that maximum valence is equal to 4. Note,
however, that the wave front condition~\eqref{msc} is only an upper
bound, and does not say whether interaction
vertices will actually contribute to $\WF(C)$ or not. 
We have checked this for the OPE-coefficient in front of 
the identity operator in the expansion of the product 
$\phi(x_1)_I \cdots \phi(x_4)_I$ of four interacting fields, 
to first order in perturbation theory
in Minkowski space, where a contribution 
from an interaction vertex in $\Gamma_{4,1}$
would be allowed according to the above 
estimate~\eqref{msc}. Using our definition~\eqref{cijk} of $C$ 
and using the integrals in~\cite{scharfetal}, we found in this example that 
such a contribution is actually absent from $\WF(C)$. 
Hence, the estimate~\eqref{msc} is not sharp.

Let us now prove the microlocal condition~\eqref{msc}. 
By eq.~\eqref{cijk}, the
microlocal spectrum condition will follow if we can show that 
if 
\ben
u_n(x_1, \dots, x_n) = 
\Psi^{j} 
\left( \prod_{r=1}^n \R_{n_r}(\O_{i_r}(x_r); I^{\otimes n_r}) \right) \, ,   
\een 
then $\WF(u_n) \subset \cup_m \Gamma_{n,m}$. To prove this
statement, we expand the retarded products as in eq.~\eqref{rexp}, then 
multiply them using the Wick expansion formula~\eqref{89}, and finally 
apply the functional $\Psi^j$. The result will be sum of expressions
each of which is a product of $r$'s, of $H$'s and expectation values 
in $\Psi^j$ of locally normal ordered expressions, which are integrated 
over interaction vertices against the smooth test function $\chi$ of
compact support appearing as infrared cutoff in the Lagrange density, $\L$. 
The expectation value in $\Psi^j$ of any locally normal ordered
expression is smooth, the wave front set of the $r$'s is given above
in (r6), see~\eqref{wfr}, while the wave front property of $H$ is 
$\WF(H)=\{(y_1, k_1; y_2, k_2); k_1 = p,
k_2 = -p, \,\, p \in \partial V_+^*\}$, where $p$ is a coparallel,
cotangent vector field along a null geodesic (edge) joining $y_1,
y_2$. We now combine these facts using the wave-front set calculus of
H\" ormander, by which we mean the following theorems about the 
behavior of the wave front set under the operations of smoothing, 
and products~\cite{hormander}:
Let $X,Y$ be manifolds (in our applications, they are Cartesian powers
of $M$). If $K \in \D'(X\times Y)$ is a distribution and $f \in
\D(Y)$ a smooth test function, then the distribution $u(x) = \int_Y
K(x,y) f(y) d\mu(y)$ has wave front set 
\ben\label{Ku}
\WF\left( u \right) \subset \{(x,k) \in T^*X; \,\, (x,k;y,0) \in \WF(K) \}
\een
Secondly, let $u,v \in \D'(X)$ so that $[\WF(u) + \WF(v)] \cap \{0\} =
\emptyset$. Then the distributional product $uv$ is defined and has
wave front set
\ben\label{uv}
\WF(uv) \subset \{(x,k+p) \in T^*X; \,\, (x,k) \in \WF(u) \cup \{0\},
\,\, (x,p) \in \WF(v) \cup \{0\} \} \, .
\een
Applying these rules to the above products of $r$'s and $H$'s, 
we essentially obtain that $\WF(u_n)$ is a subset of $\cup_m \Gamma_{n,m}$. 
For example, the momentum conservation 
rule in the third line of eq.~\eqref{gamtdef} follows from 
the additive and smoothing properties~\eqref{Ku},~\eqref{uv} 
combined with the fact that we 
are integrating the interaction vertices against the smooth
testfunction, $\chi$, of compact support. Similarly, we obtain the second line 
from the additive property~\eqref{uv}. Finally, we need to prove the 
support restriction on the interaction vertices in the fourth line of~\eqref{gamtdef}. 
For this, we note that the contribution to the wave front set of $u_n$ 
from the interaction vertices $y_k$ arises only from points  
that are in the support of the interaction, $I$, i.e., 
in the support of $\chi$. Let $U$ be an arbitrary small neighborhood
of  $J^-(\{x_1, \dots, x_n\}) \cap J^+(\{x_1, \dots, x_n\})$, and let 
$\chi'$ be a cutoff function supported in $U$. Then $\chi-\chi'$
is supported outside of the domain of dependence $D(\{x_1, \dots,
x_n\})$, and so there exists by~\eqref{vov} a unitary $V$ such that 
\ben
\prod_k \O_{i_k}(x_k)_{I'} 
= V \left\{ \prod_k \O_{i_k}(x_k)_{I} 
\right\} V^* \, . 
\een
Thus, because of~\eqref{Cdef}, we see that changing $\chi$ to
$\chi$ is equivalent to changing the standard functionals 
from $\Psi^j(\dots)$ to $\Psi^j(V \dots V^*)$. 
We claim that this would not, however, change our above wave 
front argument. Indeed, the only property of the functionals 
that was used in the above wave front set argument was that 
the expectation values of locally normal ordered expressions in 
$\Psi^j$ are smooth. This does not change if we change the standard functionals 
from $\Psi^j(\dots)$ to $\Psi^j(V \dots V^*)$. Consequently, 
we have shown that contributions to~\eqref{gamtdef} arise only from 
interaction vertices $y_k$ in $U$. Since $U$ was an arbitrarily small 
neighborhood of $J^-(\{x_1, \dots, x_n\}) \cap J^+(\{x_1, \dots,
x_n\})$, the support restriction in the last line of~\eqref{gamtdef}
follows.

\medskip

We next show that the OPE coefficients have the following local and 
covariance property: Let $f: M \to M'$ be a causality preserving
isometric embedding, let $C_I$ respectively $C'_{I'}$ be the OPE
coefficients on the respective spacetimes, 
and let $\delta > 0$ be given. Finally, assume that there are
open neighborhoods $U \subset M$ and $U' \subset M'$ with $f(U)
\subset U'$ where the cutoff functions 
$\chi$ respectively $\chi'$ implicit in the interactions 
$I$ and $I'$ are equal to 1.
Then, supposing 
that $\Delta$ in eq.~\eqref{ansatz} is chosen as in Theorem 1, 
we have on $U$
\ben
\label{Cloc}
f^* C'_{I'} \sim_{\T, \delta} C^{}_I \quad \text{for all trees $\T$.}
\een
This condition essentially follows from the fact that the interacting 
fields are local and covariant, in the sense of~\eqref{loc}, which
follows in turn from the fact that the individual terms in the
perturbation expansion of the interacting fields are local and
covariant. However, a complication arises from the fact that 
the algebra embedding $\alpha_f$ in~\eqref{loc} is not simply given 
in terms of the corresponding free field homomorphism, but is more 
complicated~\cite{hw3}. 

Instead of taking into account the more complicated definition 
of $\alpha_f$ at the interacting level, one can also more directly 
prove~\eqref{Cloc}. For this, we note that, if the cutoff function 
$\chi'$ on $M'$ were such that $f^* \chi' = \chi$, then we would 
have equality in~\eqref{Cloc}, because the retarded products and
standard functionals which are the ingredients in 
the definition of $C_I$ have a local and covariant dependence 
simultaneously on {\em both} $\chi$ and the metric $g$ implicit in
$I$, by property (r5). Thus, it is sufficient to show
that $C_I$ is essentially independent of the cutoff function $\chi$. In 
other words, if $\chi$ and $\chi'$ are two cutoff functions (on the
{\em same} spacetime) which are equal to 1 on $U$, 
and if $C_I$ and $C_{I'}$ are the corresponding OPE 
coefficients, then we must show that
\ben\label{Ciip}
C_I \sim_{\delta, \T} C_{I'}
\een
holds on $U$.
To prove this statement, we simply apply the 
functionals $\Psi(V \dots V^*)_I$ 
to the remainder
$\N_{I'}$ of the OPE formed using the coefficients $C_{I'}$, 
where $V$ is the unitary in eq.~\eqref{vov} relating the 
interactions $I$ and $I'$. Then we find $\Psi( V \N_{I'} V^*)_I = C_{I}-C_{I'}$.
Since $\N_{I'}$ is the remainder of the OPE, and since $\Psi(V \dots V^*)_I$ is a Hadamard
functional, it follows that $\epsilon^{-\delta}(C_I-C_{I'}) \circ
\psi_\T(\epsilon)$ will go to zero by theorem~1, 
which is what we needed to show.

\medskip

In~\cite{hw3}, it was shown that the perturbative interacting fields
obey a ``local covariant renormalization group flow''. The
construction of this flow involves the consideration of a 
1-parameter family $\lambda^2 g$ of conformally rescaled metrics, 
where $\lambda \in \mr_{>0}$, and states how the interacting fields
$\O_i(x)_I$ change under such a rescaling. In~\cite{hw5}, a simple
general argument was given 
that the existence of such a local covariant renormalization group
flow implies a corresponding flow of the OPE coefficients if the
theory possesses an OPE in the sense described in the previous
section. The key assumption on the nature of the RG made in~\cite{hw5}
was that there exists a suitable ``basis'' of functionals which 
is in some sense ``dual'' to the fields. This is the case in
perturbation theory, on account of~\eqref{ansatz}. Hence, it follows
by the argument of~\cite{hw5} that 
\ben
Z(\lambda)^{i_1}{}_{j_1} \cdots Z(\lambda)^{i_1}{}_{j_1} [^t
Z(\lambda)^{-1}]_k{}^l
C_{i_1 \dots i_n}{}^k[M,g]_I \sim_{\T,\delta} 
C_{j_1 \dots j_n}{}^l[M,\lambda^2 g]_{I(\lambda)}
\,  
\een
for all trees $\T$. Here, the prefactors are  
linear maps (whose construction and properties was described
in~\cite{hw3}) 
\ben
Z(\lambda)^{i}{}_j \in {\rm End}(\E_j, \E_i) \, , 
\een 
where $\E_i$ is the vector bundle in which the field $\O_i$ lives, and 
$I(\lambda)$ is an interaction of the same form as $I$ with suitable 
``running'' couplings $\kappa_i(\lambda)$, 
whose construction was also described in~\cite{hw3}. 

\medskip

An associativity property for the OPE
coefficients may be formulated as follows (see~\cite{hw5} for
details). Let $\x(\epsilon)$ be a curve in configuration space 
representing the merger of the points according to 
a tree $\T$, that is, $\psi_\T(\epsilon): \x \mapsto \x(\epsilon)$. 
In this situation, we should be able to perform the OPE successively, 
in the hierarchical order represented by the tree, thus
leading to some kind of ``asymptotic factorization''. That is, 
we should be allowed to first perform the OPE for each subtree, and 
then successively the OPE's corresponding to the branches relating the
subtrees, and so fourth. For example, for the tree $\T$ given in the 
figure on p.~16, we should be allowed to perform the OPE successively 
as indicated by the brackets $(\O_1 \O_2)(\O_3 \O_4)$.
To formulate this condition more precisely, 
we recall the notation $s(e), t(e)$ for the 
source and targets of an edge, $e$, in the tree $\T$. Furthermore, for 
$\x = (x_1, \dots, x_n) \in M^n$, and  
for each node $S \in \T$ of the tree, let us set
\ben
x_S = x_{m(S)}, \quad m(S) = {\rm max} \{i: \, i \in S\} \, .
\een
Finally, we consider maps $\vec i: \T \to \I$ which associates with 
every node $S \in \T$ of the tree an element $i_S \in \I$, the index 
set labelling the fields. With these notations in place, the 
associativity property can be stated as
follows. Let $\delta > 0$, let $\T$ be a tree, and let $M_S^n$
be the set of all ``spacelike configurations'' $\x = (x_1, \dots, x_n)$ 
\ben
M^n_S = \{ \x \in M^n_0; \quad x_i \notin J^+(x_j) \cup J^-(x_j)\, \, 
\text{forall $i,j$}\} \, .
\een
Then, on $M_S^n$, we have
\begin{multline}
\label{41}
C_{i_1 \dots i_n}{}^j(x_1, \dots, x_n)_I \\
\sim_{\T, \delta} 
\sum_{\vec i} \prod_{S \in \T} 
C_{ \{i_{t(e)}; \text{$e$ such that $s(e) = S$}\} }{}^{i_{S}}
\left( \big\{ x_{t(e)}; \text{$e$ such that $s(e)=S$} \} \right)_I \, , 
\end{multline}
where the sum is over all $\vec i$, with the properties that 
\bena
&&i_{\{k\}} = i_k, \quad k=1, \dots, n, \quad i_{\{1, \dots, n\}} = j, \\
&&\sum_{e: s(e) = S} [i_{t(e)}] < \delta_S \quad \forall S \in \T \, , 
\eena
where $\delta_S>0$ are chosen sufficiently large. 
Note that it makes sense to consider the relation $\sim_{\delta, \T}$ with respect to the
open subset of spacelike configurations, because 
a configuration remains spacelike when scaled down by 
$\psi_\T(\epsilon)$, at least provided the points $\x \in M^n_S$ are in a 
sufficiently small neighborhood of the total diagonal, which we assume
is the case. The technical reason 
for restricting the OPE to pairwise spacelike related points
is that the OPE coefficients $C$ are smooth on $M_S^n$, by the microlocal 
spectrum condition (see eq.~\eqref{msc}), 
and so convergence in the sense of $\sim_{\delta, \T}$
is more straightforward to study. Furthermore, since the 
interacting fields commute for spacelike related points, 
there are no ordering issues when working with the configurations 
in $M_S^n$. Also, from a physical viewpoint, 
the notion of ``short distances'' is somewhat unclear if lightlike 
directions are included. 

In~\cite{hw5}, it is shown that associativity in the above sense is 
an automatic consequence if the theory also possess a suitable local
covariant renormalization group with suitable properties, and if the 
OPE holds not only for each fixed spacetime and fixed choice of 
couplings, but instead also uniformly
in a suitable sense for smooth families of metrics and couplings 
(termed ``condition (L)'' in that paper). As we have already described, 
the existence of a local covariant 
renormalization group in perturbation theory was established
in~\cite{hw3}. It is a general property of the perturbative 
renormalization group that 
$Z^i{}_j(\lambda)$ is given by $\lambda^{-[i]}$ times a 
polynomial in $\ln \lambda$ at each finite order in perturbation 
theory~\cite{hw3}, and that the running couplings $\kappa_i(\lambda)$
in $I(\lambda)$ have a power law dependence $\lambda^{4-[i]}$, 
which is modified by 
polynomials in $\ln \lambda$ at any given order in perturbation 
theory. These are essentially the properties for the renormalization 
group required in (L) at the perturbative level, except that the 
running of couplings $I(\lambda)$
is not exactly smooth at $\lambda = 0$ as required in (L), 
but instead contain logarithmic terms at each order in perturbation theory. 
However, the argument given in~\cite{hw5} is insensitive to 
such logarithmic corrections. To also establish the desired smooth 
dependence of the OPE-coefficients under smooth variations of 
the metric required in (L) at the perturbative level, 
it is necessary to go through the proof of 
Theorem 1 for families of spacetimes and corresponding 
families of states depending smoothly on a parameter 
in the sense~\cite{hw2} and analyze the 
behavior of the constructions under variations of the parameter. 
This can indeed be done, 
using the smooth dependence of the retarded products under such
parameters~\cite{hw2}, as well as the techniques and type of arguments employed in the
appendix of~\cite{hw4}. However, even though the repetition of these 
arguments is in principle straightforward, that analysis is
quite lengthy and cumbersome, and not very illuminating. It is
therefore omitted. Since a perturbative version of 
condition (L) holds, Theorem 1 of~\cite{hw5} then implies that the
associativity property holds on the space $M_S^n$ of pairwise
spacelike configurations. 

\medskip

In the remainder of this section we will 
prove that the OPE coefficients themselves can be expanded for
asymptotically small distances in terms of curvature terms and 
Minkowski distributions in the tangent space (for spacelike 
related configurations in $M^n_S$, to which we shall restrict
ourselves in the remainder of this section).
The construction of this ``scaling expansion'' involves 
the Mellin transform, $\M[f, z]$, of a function $f(x)$ defined 
on $\mr_{>0}$ vanishing near infinity, with at most polynomial type 
singularity~\cite{mellin} as $x \to 0$. It 
is defined by 
\ben
\M[f, z] = \int_0^\infty x^{\i z-1} f(x) \, dx \, ,
\een
and is an analytic function of $z$ for sufficiently small ${\rm Im}(z)
< y_0$ 
where $y_0$ depends upon the strength of the singularity
of $f$. The inverse Mellin transform of $F(z) = M[f, z]$ is given by
\ben
\M^{-1}[F, x] = \frac{1}{2\pi \i} 
\int_{-\i \infty + c}^{+\i \infty + c} x^z F(z) \, dz
\, 
\een
where the integration contour is to the right of all poles of $F(z)$
in the complex $z$-plane. The Mellin transform is useful 
in the context of functions $f(x)$ possessing 
near $x=0$ an asymptotic expansion of the form
\ben\label{fexp}
f(x) \sim \sum_p \sum_l a_{p, l} x^{-p} (\ln x)^l \, ,  
\een 
where $p$ is bounded from above, and where the sum over $l$
is finite for any $p$.
It can be seen that the Mellin transform of such a function 
possesses isolated poles at $z = \i p$ in the complex plane, with 
finite multiplicities. Furthermore, the 
asymptotic expansion coefficients 
$a_{p, l}$ are the residues of the Mellin transform, 
i.e., 
\ben
a_{p, l} = \frac{1}{l!} {\rm Res}_{z=-\i p} 
\Bigg\{
(z+ \i p)^l \M[f, z] \Bigg\} \, .
\een
We now define, for each tree $\T$, distributions $C^\T_I$ that give
the desired scaling expansion of the OPE coefficient $C_I$ relative 
to the scaling function $\psi_\T(\epsilon): M^n_0 \to M^n_0, \x
\mapsto \x(\epsilon)$ 
defined above in eq.~\eqref{psidef}, by extracting the poles of $C_I \circ
\psi_\T(\epsilon)$ in $\epsilon$ using the Mellin transform. In order to 
do this, let $\x = (x_1, \dots, x_n) \in M^n_S$ be a spacelike configuration of $n$ points, let $\T$ 
be a tree, and let
$\epsilon \mapsto \x(\epsilon)$ 
be the corresponding curve in $M^n_S$. By the microlocal spectrum property, 
the OPE coefficients are 
smooth on $M^n_S$, so we may consider $[C_I \circ \psi_\T(\epsilon)](\x)
= C_I(\x(\epsilon))$ as a 
smooth function in $\epsilon$ at any fixed value of the argument 
$\x$. As we will show in the proof of the next theorem, 
if we fix the parameter $\delta>0$
in the operator product expansion, this function has an expansion 
of the form
\ben\label{polyhom}
C_I(\x(\epsilon)) = \sum_{p} \sum_{l} a_{p, l}(\x) 
\epsilon^{-p} (\ln \epsilon)^l + \dots \, ,
\een
near $\epsilon=0$,
where the dots stand for a remainder 
vanishing faster than $\epsilon^\delta$. Here, $p$ is 
in the range from $-\delta$ to $D={\rm depth}(\T) \cdot 
(-[k] + \sum_j [i_j])$, and the sum over $l$ is finite 
for each $p$, at any given order in perturbation theory\footnote{Note
  however that the range of $l$ increases with the perturbation order.}.  
Consequently, we can define the Mellin transform\footnote{To make this
expression well defined, we need to arbitrarily cut off the integral for large
$\epsilon$ (where the map $\psi_\T(\epsilon)$ is not well-defined anyway). How
this cutoff is chosen does not affect the following discussion.} of this function 
in the variable $\epsilon$
\ben
\label{mellin}
\M^\T(\x, z) \equiv \M[C_I \circ \psi_\T(\epsilon), z] 
= \int_0^\infty C_I (\x(\epsilon)) \, 
\epsilon^{\i z - 1} \, d\epsilon \, , 
\een
which is now a function of $\x \in M^n_S$ that is analytic in 
$z \in \mc$ for sufficiently small ${\rm Im}(z)$. Furthermore, 
by the above expansion~\eqref{polyhom}, it is meromorphic on a domain
including ${\rm Im}(z) \le \delta$, with poles possibly at 
$\i \delta, \i(\delta -1), \i(\delta-2),
\dots, -\i D$. Let us now choose a contour $\mathcal C$ around these points
as illustrated in the figure.  

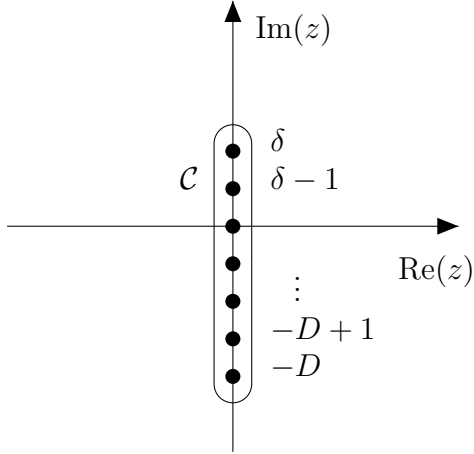
\setlength{\unitlength}{1cm}
\begin{center}
\begin{picture}(6,6)(0.0,0.0)
\put(0,3){\vector(6,0){6}}
\put(5.2,2.3){${\rm Re}(z)$}
\put(2.3,3.5){$\mathcal C$}
\put(3,0){\vector(0,6){6}}
\put(3,2.5){\oval[0.5](0.5,3.7)}
\put(3.3,5.5){${\rm Im}(z)$}
\put(3,1){\circle*{0.2}}
\put(3.5,1){$-D$}
\put(3,1.5){\circle*{0.2}}
\put(3.5,1.5){$-D+1$}
\put(3,2){\circle*{0.2}}
\put(3,2.5){\circle*{0.2}}
\put(3,3){\circle*{0.2}}
\put(3,3.5){\circle*{0.2}}
\put(3.5,3.5){$\delta-1$}
\put(3,4){\circle*{0.2}}
\put(3.5,4){$\delta$}
\put(3.8,2){$\vdots$}
\end{picture}
\end{center}

\noindent
Define
\ben
\label{Ctdef}
C^\T(\x)_I \equiv 
\frac{1}{2\pi \i} 
\oint_{{\mathcal C}} \M^\T(\x, z)
\, dz \, .  
\een
Concerning this function on $M_S^n$, we have the following theorem.
\begin{thm}
\begin{enumerate}
\item We have $C^\T_I \sim_{\T, \delta} C_I$ for spacelike
  configurations, and therefore
\ben\label{opeT}
\O_{i_1}(x_1)_I \O_{i_2}(x_2)_I \cdots \O_{i_n}(x_n)_I 
\sim_{\T,\delta} \sum_{[k] \le \Delta} 
C^\T_{i_1 i_2 \dots i_n}{}^k(x_1, x_2, \dots, x_n)_I \,  
\O_{k}(x_n)_I \,  
\een  
on $M_S^n$.
\item $C^\T_I$ is local and covariant, i.e., if $f:(M,g) \to (M',g')$ is 
an orientation, causality preserving 
isometric embedding then $C^\T_I = f^* C^{\prime \T}_{I'}$. In particular, 
$C^\T_I$ does not depend on the choice of cutoff function $\chi$ used in the 
definition of the interacting field.
\item The expression $C^\T_I(\x)$ is the sum of residue of $\M^\T(z,\x)$
corresponding to the poles in the contour ${\mathcal C}$ 
in~\eqref{Ctdef}, 
\ben\label{scaling0}
C^\T_I(\x) = \sum_{p \ge -\delta} {\rm Res}_{z=-\i p} \left\{ \M^\T(\x,
  z) \right\} \, .
\een
These have the following form: 
\ben
\label{scaling}
{\rm Res}_{z= -\i p} \left\{ \M^\T(\x, z) \right\}
= \sum_a P_a[\nabla_{(\alpha_1} \cdots \nabla_{\alpha_k)}
R_{\mu_1\mu_2\mu_3\mu_4}(x_n)] \, 
W^a(\xi_1, \dots, \xi_{n-1}) \,  , 
\een
where $P_a$ is a polynomial in the Riemann tensor and (finitely many)
of its covariant derivatives evaluated at $x_n$, valued in some 
tensor power of the tangent space $(T_{x_n} M)^{\otimes a}$, 
while $\xi_i$ are the Riemannian normal coordinates of $x_1, \dots,
x_{n-1}$ around $x_n$, identified with vectors in 
$\mr^4$ via a tetrad. The sum over $a$ is finite,
and each $W_a$ is a Lorentz covariant distribution on 
$\mr^{4(n-1)}$ (defined on spacelike configurations), 
that is valued in $(\mr^4)^{\otimes a}$ 
(identified with $(T^*_{x_n} M)^{\otimes a}$ via the tetrad),
depending polynomially on $m^2, \alpha$. Thus, 
for any proper, orthochronous Lorentz transformation $\Lambda$, 
we have 
\ben\label{lorentz}
W_a(\Lambda \xi_1, \dots, \Lambda \xi_{n-1}) = D^b{}_a (\Lambda)
W_b(\xi_1, \dots, \xi_{n-1}) \, , 
\een
where $D(\Lambda)$ is the corresponding tensor representation.

\item There are Lorentz invariant distributions $V_{a, l}$ such that 
\ben
W_a(\vec \xi(\epsilon)) = \epsilon^{N} \left[ W_{a}(\vec \xi)
+ \sum_l (\ln \epsilon)^l V_{a,l}(\vec \xi) \right] \, , 
\een
where the sum is only over a finite range of $l$, at any given, but 
finite order in perturbation theory.
\end{enumerate}
\end{thm}
\paragraph{Remarks:} 1) Equations~\eqref{scaling0},~\eqref{scaling} 
constitute the claimed scaling expansion. That scaling expansion is
similar in nature to a corresponding expansion derived in~\cite{hw2}
for the short distance behavior of time ordered products. However,
note that the scaling expansion above is more general than that
derived in~\cite{hw2}, because it
involves the consideration of more general scaling functions
$\psi_\T(\epsilon)$ corresponding to general trees. Also, the
Mellin transformation technique was not used in~\cite{hw2}.

2) The restriction to spacelike configurations has mainly been made 
for technical reasons, to avoid technical issues that could arise when 
taking the Mellin transformation of a distribution. However, we expect 
that all constructions and properties summarized in the above theorem 
hold for all configurations, i.e., in the sense of distributions on $M^n$.

\medskip
\noindent

{\em Proof}: Let us first argue that the 
claimed meromorphicity of the Mellin transform of
$C(\x(\epsilon))_I$ holds, or equivalently, that it 
has an asymptotic expansion as claimed in eq.~\eqref{polyhom}. 
For this, we
first consider the simple tree $\T=\{S_0, S_1, 
\dots, S_n\}$ with one root $S_0=\{1, \dots, n\}$ and 
$n$ leaves $S_i = \{i\}$. The depth of this tree is $depth(\T)=1$, and 
$\psi_\T(\epsilon)$ is simply the map which multiplies the 
Riemann normal coordinates of $x_i$ relative to $x_n$ by $\epsilon$.
To analyze the corresponding scaling of the OPE coefficients, 
we note that, by the local and covariance property of the OPE 
coefficients, a rescaling of the arguments $x_i$ is equivalent
to changing the metric from $g$ to $s_\epsilon^* g$, where 
$s_\epsilon: M \to M$ is the diffeomorphism that scales the Riemannian
coordinates of a point around $x_n$ by $\epsilon$. Thus, we have 
have 
\ben
C_I[M,g] \circ \psi_\T(\epsilon) \sim_{\T, \delta} C_I[M,
s_\epsilon^* g] \, 
\een
for this tree.
Next, we use the fact that in perturbation theory, there 
exists a local and covariant renormalization group~\cite{hw3}. This implies
that, up to terms of order $\epsilon^\delta$ 
\ben\label{000}
Z(\epsilon)^{i_1}{}_{j_1} \cdots Z(\epsilon)^{i_1}{}_{j_1} [^t
Z(\epsilon)^{-1}]_k{}^l
C_{i_1 \dots i_n}{}^k[M,g(\epsilon)]_{I(\epsilon)} = 
C_{j_1 \dots j_n}{}^l[M,g]_{I} \circ \psi_\T(\epsilon)
\,  , 
\een
where $g(\epsilon) = \epsilon^{2} s_\epsilon^* g$. The point is 
now that the metric $g(\epsilon)$ has a smooth dependence upon
$\epsilon$, as may be seen by rewriting it in Riemannian normal 
coordinates as $g_{\mu\nu}(\epsilon \xi) d\xi^\mu d\xi^\nu$. The OPE
coefficients in turn have a smooth dependence upon smooth variations of the 
metric, since all the quantities in their definition have this 
property~\cite{hw2}. The only singular terms (in $\epsilon$) 
in the expression on 
the left side can therefore come from (a) the running 
couplings in $I(\epsilon)$ and (b) the $Z(\epsilon)$-factors. 
However, by the general analysis of the local renormalization 
given in~\cite{hw3}, these quantities are polynomials 
in $\epsilon^{-p} (\ln \epsilon)^l$ of finite degree to any 
given order in perturbation theory. Thus, 
up to terms vanishing faster than $\epsilon^\delta$, the quantity
$C_I(\x(\epsilon))$ has an expansion of the type~\eqref{polyhom}, 
as desired. 

The argument just given may be generalized to arbitrary trees 
by an induction in the depth of the tree. For trees of depth one 
we have just proven the statement. Let us inductively assume 
that we have proven~\eqref{polyhom} for trees of depth $d$. 
To deal with trees of depth $d+1$, one notices that a 
tree of depth $\ge 2$ can always be decomposed 
into a tree $\mathcal S$ of depth one connected to the root, and 
trees $\T_1, \dots, \T_r$ attached to the leaves of $\mathcal
S$. Thus, we may write 
\ben
\T = {\mathcal S} \cup \bigcup_{t=1}^r \T_t \, , 
\een
see the figure for an example with $r=2$. 

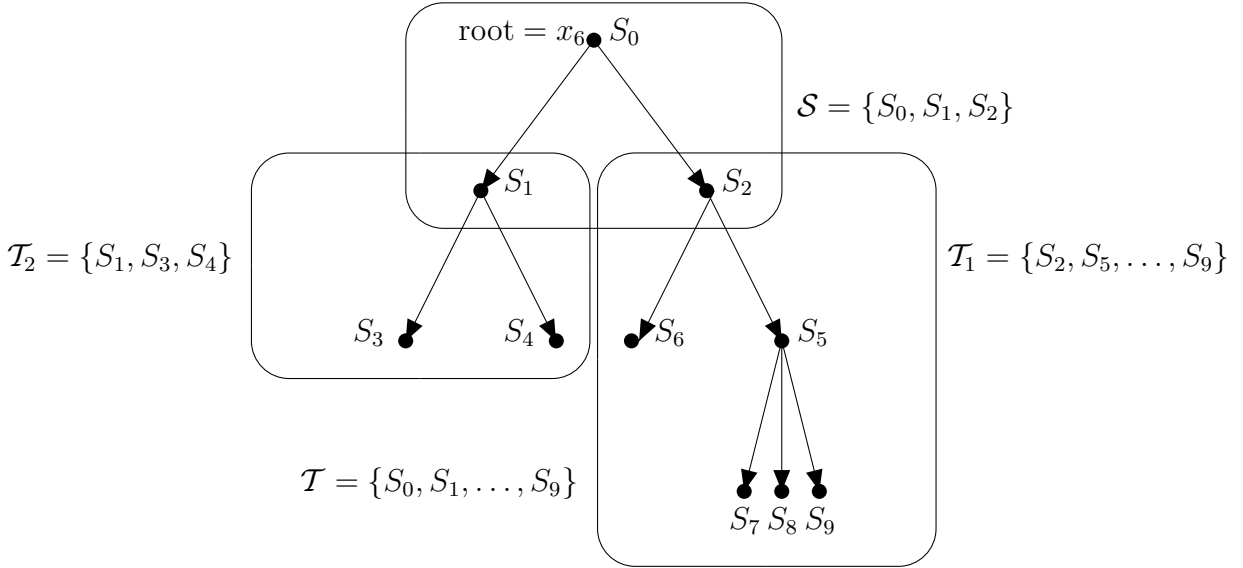
\setlength{\unitlength}{1cm}
\begin{center}
\begin{picture}(6,8)(0.0,-1)
\put(3,5){\oval[0.5](5,3)}
\put(5.7,5){${\cal S}=\{S_0, S_1, S_2\}$}
\put(5.3,1.75){\oval[0.5](4.5,5.5)}
\put(7.7,3){${\cal T}_1=\{S_2, S_5, \dots, S_9\}$}
\put(0.7,3){\oval[0.5](4.5,3)}
\put(-4.8,3){${\cal T}_2=\{S_1, S_3, S_4\}$}
\put(3,6){\circle*{0.2}}
\put(3.2,6){$S_0$}
\put(1.2,6){${\rm root}=x_6$}
\put(1.5,4){\circle*{0.2}}
\put(1.8,4){$S_1$}
\put(4.5,4){\circle*{0.2}}
\put(4.7,4){$S_2$}
\put(0.5,2){\circle*{0.2}}
\put(-0.2,2){$S_3$}
\put(2.5,2){\circle*{0.2}}
\put(1.8,2){$S_4$}
\put(5.5,2){\circle*{0.2}}
\put(5.7,2){$S_5$}
\put(3.5,2){\circle*{0.2}}
\put(3.8,2){$S_6$}
\put(3,6){\vector(-1.5,-2){1.5}}
\put(3,6){\vector(1.5,-2){1.5}}
\put(1.5,4){\vector(-1,-2){1}}
\put(1.5,4){\vector(1,-2){1}}
\put(4.5,4){\vector(1,-2){1}}
\put(4.6,4){\vector(-1,-2){1}}
\put(5.5,2){\vector(-0.5,-2){0.5}}
\put(5.5,2){\vector(0,-2){2}}
\put(5.5,2){\vector(0.5,-2){0.5}}
\put(5,0){\circle*{0.2}}
\put(5.5,0){\circle*{0.2}}
\put(6,0){\circle*{0.2}}
\put(4.8,-0.5){$S_7$}
\put(5.3,-0.5){$S_8$}
\put(5.8,-0.5){$S_9$}
\put(-0.9,0){${\mathcal T} = \{S_0, S_1, \dots, S_9\}$}
\end{picture}
\end{center}

The key point is now that the map 
$\psi_\T(\epsilon)$ factorizes by the inductive nature of its
definition~\eqref{psidef}, while the OPE coefficient $C_I$
factorizes by the associativity property~\eqref{41}
under this decomposition of the tree. This gives
\ben\label{recursion}
C_{i_1 \dots i_n}{}^k \circ \psi_\T(\epsilon) \sim_{\T, \delta} 
\left[ 
C_{j_1 \dots j_r}{}^k \cdot \bigotimes_{t=1}^r
C_{\{ i_S; \,\, S \in {\rm Leaves}(\T_t)\}}{}^{j_t} \circ \psi_{\T_t}(\epsilon)
\right] \circ \psi_\S(\epsilon) \, , 
\een
where we note that, since $\S$ has depth one, the map
$\psi_\S(\epsilon)$ is given in terms of the diffeomorphism
$s_\epsilon:M \to M$ which scales the Riemann
normal coordinates around $x_n$ of points by $\epsilon$. 
On the right side of this expression, we can now 
apply the induction hypothesis to the expression in brackets, because each of the trees $\T_t$
has depth $\le d$. Furthermore, since $\psi_\S(\epsilon)$ is given 
by the diffeomorphism $s_\epsilon$, we may again 
apply the general covariance and renormalization group property 
as we did above to convert the action of $s_\epsilon$ into 
a smooth change $g(\epsilon)$ in the metric and $Z(\epsilon)$-factors
depending polynomially on $\epsilon^{-p} (\ln \epsilon)^l$. 
This proves the equation~\eqref{polyhom}. 

\medskip
We next come to the actual proof of the theorem.
To prove 1), we need to show that $ \epsilon^{-\delta} \, 
(C^{}_I-C^\T_I) \circ \psi_\T(\epsilon) \to 0$ as $\epsilon \to 0$, 
to any finite but arbitrary order in perturbation theory. 
We have, using the definition of $C^\T_I$ and the relation 
$\psi_\T(\epsilon) \circ \psi_\T(\epsilon') = \psi_\T(\epsilon\epsilon')$
\bena
\label{125}
C^\T_I \circ \psi_\T(\epsilon) &=& \sum_{p \ge -\delta} 
{\rm Res}_{z = -\i p} \int_0^\infty
(C_I \circ \psi_\T(\epsilon')) \circ \psi_\T(\epsilon) \,  
\epsilon^{\prime \i z -1} \, d\epsilon' \nonumber\\
&=& \sum_{p \ge -\delta} {\rm Res}_{z= -\i p} \Bigg\{ 
{\rm e}^{\i z \ln \epsilon} \, \M^\T(\x, z)
\Bigg\} \nonumber\\
&=& \sum_{p \ge -\delta} 
\epsilon^{-p} \sum_l \frac{(\ln \epsilon)^l}{l!} {\rm Res}_{z= -\i p}
    \left\{ (z+\i p)^l \M^\T(\x, z) \right\} \nonumber\\
&=& \sum_{p \ge -\delta} \sum_l \epsilon^{-p} 
(\ln \epsilon)^l a_{p, l}(\x) \, , 
\eena
where we have performed a change of integration variables in 
the second step. Comparing with~\eqref{polyhom}, this formula 
implies that $C^\T(\x(\epsilon))_I$ differs from $C(\x(\epsilon))_I$ 
by a term vanishing faster than $\epsilon^\delta$. 
This proves the assertion 1). 

To prove 2), we recall that we have already proven above that, 
at each order in perturbation theory, 
$f^* C_{I'}[M',g'] \sim_{\delta, \T} C_I[M,g]$, so the 
difference between these two terms vanishes faster than 
$\epsilon^\delta$. This difference 
term will change the Mellin transform $\M^\T(z)$ only by 
a term that is analytic in a domain including ${\rm Im}(z) \le \delta$, and 
thus will not contribute to $C^\T_I[M,g]$ respectively 
$C^\T_{I'}[M',g']$, because the contour integral of a holomorphic 
function vanishes. 

For 3), consider again the metric $g(\epsilon) = \epsilon^2
s_\epsilon^* g$. Its components in Riemann normal coordinates around $x_n$
have a Taylor expansion of the form 
$g_{\mu\nu}(\epsilon \xi) = \sum \epsilon^N P_N(\nabla^k 
R_{\alpha_1\alpha_2\alpha_3\alpha_4}(x_n), \xi^\rho)_{\mu\nu}$, 
where $P_N$ are polynomials. Since the $C^\T_I$ are local and
covariant by 2), it follows
that they can be viewed as functionals of 
the Riemann tensor and its covariant derivatives at point $x$
(or $\xi = 0$) which enters via $P_N$. Thus, it  
follows that the $N$-th $\epsilon$-derivative is
\ben
\frac{\partial^N}{\partial \epsilon^N} 
C^\T_I[g(\epsilon)](\x) |_{\epsilon = 0}
= \sum_{N_1+\dots+N_r=N} W^{N_1 \dots N_r}(\vec \xi) \prod_i
P_{N_i}[\nabla_{(\alpha_1} \cdots \nabla_{\alpha_k)}
R_{\mu_1\mu_2\mu_3\mu_4}(x_n), \xi^\nu_j] \, 
\, , 
\een 
where
\ben
W^{N_1 \dots N_r} = \frac{\partial^r C^\T_I}{\partial P_{N_1} \cdots
  \partial P_{N_r}} \bigg|_{g = \eta} \, .
\een
We define $W^a$ and $P_a$ by the above relation, i.e., 
$P_a$ is the appropriate product of the $P_{N_i}$, with the 
polynomial $\xi_i$-dependence taken out and absorbed in the 
definition of $W^a$. 
To prove the desired relation~\eqref{scaling}, we must now show that the
$\epsilon$-derivatives of $C^\T_I[g(\epsilon)]$ at $\epsilon=0$ vanish for $N$
sufficiently large. This quantity arises from the quantity 
$C_I[g(\epsilon)] \circ \psi_\T(\epsilon')$ by taking a Mellin 
transform in $\epsilon'$, then taking $N$ derivatives with respect to 
$\epsilon$ at $\epsilon=0$, and finally extracting the residue in $z$. It follows that 
we only need to show that the quantity obtained by taking 
$N$ derivatives with respect to $\epsilon$ of $C_I[g(\epsilon)] \circ 
\psi_\T(\epsilon')$ vanishes faster than ${\epsilon'}^\delta$, because
such a term would not give rise to poles of the Mellin transform 
in the domain ${\rm Im}(z) \le \delta$. Consider first the case when 
$\T$ has depth one, so that $\psi_\T(\epsilon')$ is simply a dilation 
of the Riemann normal coordinates by $\epsilon'$. As above in
eq.~\eqref{000}, 
by combining the renormalization group and general covariance, 
the action of such a dilation may be translated into changing
$g(\epsilon)$ to $g(\epsilon \epsilon')$, along with a suitable set 
of $Z(\epsilon')$-factors, and running couplings in
$I(\epsilon')$. The point is now that the $N$ derivatives on
$\epsilon$ will produce, when acting on $g(\epsilon\epsilon')$, 
precisely $N$ positive powers of $\epsilon'$. Thus, if $N$ is
sufficiently large, then the resulting positive powers will dominate
the corresponding negative powers in the $Z(\epsilon')$-factors, and 
we get the desired result. The generalization of this argument to 
arbitrary trees $\T$ can be done via an induction argument based 
upon formula~\eqref{recursion}, similar to the one given there.

To prove the Lorentz invariance of the $W^a$, let 
us now define, for each Lorentz transformation 
$\Lambda \in {\rm SO}(3,1)^\uparrow_0$ the diffeomorphism
$f_\Lambda: \xi \mapsto \Lambda \xi$. It defines a causality and orientation 
preserving isometric embedding between the spacetimes
$(M,g)$ and $(M, f_\Lambda^*g=g_\Lambda)$ with the same orientations. 
Thus, using the local covariance property $f_\Lambda^*C^\T[g]
= C^\T[g_\Lambda]$, and the transformation property of 
$f_\Lambda^* P^a = D^a{}_b(\Lambda) P^b$ 
under this diffeomorphism, it follows that 
\begin{multline}
\sum_a P_a[\nabla^k R_{\alpha_1\alpha_2\alpha_3\alpha_4}(x_n), \xi_i^\nu] 
\, 
W^a(\Lambda \xi_1, \dots, \Lambda \xi_{n-1})\\
= \sum_{a,b} P_a[ \nabla^k
R_{\alpha_1\alpha_2\alpha_3\alpha_4}(x_n), \xi^\nu_i] 
\, D^a{}_b(\Lambda)
W^b(\xi_1, \dots, \xi_{n-1}) \, .
\end{multline}
However, since this holds for all metrics, eq.~\eqref{lorentz} follows.

Using the definition of $W^a$ just given, 
item 4) immediately follows from the fact that 
$C_I$ has an expansion of the form~\eqref{polyhom}, and that 
$C_I^\T \sim_{\delta, \T} C_I^{}$. \qed

\section{Example}

We now illustrate our general method for computing the OPE in curved 
spacetime by an example. Let us summarize again the steps needed 
in this computation. 
\begin{enumerate}
\item Fix a desired accuracy, $\delta$, of the OPE, and determine
$\Delta$ as in Theorem 1.

\item Identify the retarded products in eq.~\eqref{cijk} that are needed
to compute the desired OPE coefficient to a given order in perturbation
theory and a given accuracy $\delta$, 
and determine them, using the methods of the papers~\cite{hw1,hw2}. 

\item 
Perform a local ``Wick expansion~\eqref{rexp} of all retarded
products. In places in~\eqref{cijk} where two retarded products
are multiplied, perform the products using Wick's theorem~\eqref{89}
(with $\omega_2$ in that formula replaced by $H$). Apply the 
standard functionals~\eqref{standard0} to the resulting expressions
in the way indicated in~\eqref{cijk}. 
\end{enumerate}
This yields the desired OPE coefficient. If one is interested in the
asymptotic behavior of the OPE coefficient (up to order $\delta$ in 
$\epsilon$) under a rescaling of a point $\psi_\T(\epsilon):
\x \mapsto \x(\epsilon)$ associated with a given tree $\T$, perform 
the following step:
\begin{enumerate}
\item[4.] Take the Mellin-transform of $C_I(\x(\epsilon))$ in
  $\epsilon$ as in eq.~\eqref{mellin}, and define $C^\T_I(\x)$
to be the sum of its residue at the poles 
$\i \delta, \i(\delta -1), \dots$, as in eq.~\eqref{Ctdef}. 
The result then automatically has the form of curvature terms 
times Minkowski distributions in the
relative coordinates as described in item 2) of Theorem~2, 
and it is equivalent to $C_I$ in the sense 
that $C_I^\T \sim_{\T, \delta} C_I^{}$.
\end{enumerate}
 
\medskip

As an illustration, we now apply this method to the determine the coefficient $C_I$
in the triple product of operators 
\ben
\phi(x_1)_I \phi(x_2)_I \phi(x_3)_I = \dots + C(x_1, x_2, x_3)_I \, \phi(x_3)_I
+ \dots
\een
up to first order in perturbation theory in  
the interaction $I = \int_M \L \, d\mu = 
-\frac{1}{4!} \int_M \kappa \phi^4 \, d\mu$, and 
accuracy $\delta=0$. We then discuss the scaling expansion as in item 
2) of Theorem~2. As discussed above, we impose an infrared cutoff by 
taking $\kappa(x) = \kappa \chi(x)$ at intermediate steps, 
where $\chi$ is a smooth cutoff function, but the final answers will 
not depend on the choice of $\chi$, see eq.~\eqref{Ciip}.
The different ways of scaling the 3 points together give
rise to different limiting behavior $C_I^\T$ of $C_I$. We choose to investigate 
the most interesting case when all points are scaled together at the
same rate, i.e., under the scaling map $\psi_\T(\epsilon): 
\x \to \x(\epsilon)$, 
where $\T=\{\{1,2,3\},\{1\}, \{2\}, \{3\}\}$. If $\xi_i$ are the Riemannian 
normal coordinates of the points $x_i$ around $x_3$, 
this corresponds to $\vec \xi(\epsilon) = 
(\epsilon \xi_1, \epsilon \xi_2, \epsilon \xi_3)$ with $\xi_3=0$. 

Consider first the zeroth order perturbative contribution. 
According to eq.~\eqref{cijk}, this is given 
for a general OPE coefficient of a triple product 
of operators by
\ben
C_{i_1 i_2 i_3}{}^{j}(x_1,x_2,x_3)_{(0)} =  
\Psi^{j} (\O_{i_1}(x_1) 
\O_{i_2}(x_2) \O_{i_3}(x_3)) \, ,
\een
where the reference point for the functional $\Psi^j$ (see
eq.~\eqref{standard0}) is $x_3$
throughout. We are interested in the case $\O_{i_1} = \O_{i_2} = \O_{i_3} =
\O_j = \phi$. In order to determine the action of the functional $\Psi^j$ in this case, 
we need to perform the local Wick expansion of the product 
$\phi(x_1)\phi(x_2) \phi(x_3)$, which is given by
\ben
\phi(x_1) \phi(x_2) \phi(x_3) = \,\, 
:\phi^{\otimes 3}:_\h (x_1, x_2, x_3) + H(x_1,x_2) \phi(x_3) + \,\,
{\rm cyclic}(1,2,3)
\een
Applying now the definition of the functional $\Psi^j$
(with $\O_j$ chosen to be $\phi$, and reference point $x_3$) gives
\ben\label{ooo}
C(x_1,x_2,x_3)_{(0)} = 
H(x_1,x_2)+H(x_2,x_3)+H(x_1,x_3) \, , 
\een
for any $\delta$. 

In order to determine the representer 
$C^\T_I$ to zeroth order in perturbation theory for our choice
$\delta=0$ (see eq.~\eqref{Ctdef}), we are instructed to 
compose the above result with the map $\psi_\T(\epsilon)$, then take 
a Mellin transform in the variable $\epsilon$, and then extract the 
residue at the poles $0,-\i,-2\i$ in the complex $z$-plane via 
the contour integral~\eqref{Ctdef}. As explained above, taking a
Mellin transform and then extracting those residue is a way to compute
the corresponding coefficients of $\epsilon^0, \epsilon^{-1},
\epsilon^{-2}$ in the expansion~\eqref{polyhom} of the distribution 
$C_I \circ \psi_\T(\epsilon)$ in $\epsilon$. 
In the present simple example, it is
easier to compute the coefficients directly from eq.~\eqref{ooo}, by 
using the corresponding short distance expansions~\cite{deWitt} 
around $x_3$ of the quantities $\sigma, v_n$ 
appearing in the Hadamard parametrix $H$. 
The result is 
\begin{multline}
C^\T(x_1, x_2, x_3)_{(0)} = 
\frac{1}{2\pi^2} \sum_{i<j} \Bigg\{ \frac{1}{(\xi_i-\xi_j)^2} \\
-\frac{1}{3} \frac{R_{\mu\nu\sigma\rho} \xi_i^\mu \xi_j^\nu 
\xi_i^\sigma \xi_j^\rho
}{(\xi_i-\xi_j)^4} -\frac{1}{6} 
\frac{
R_{\mu\nu}(
\xi_i^\mu \xi_i^\nu
+\xi_i^\mu \xi_j^\nu
+\xi_j^\mu \xi_j^\nu)
}{(\xi_i-\xi_j)^2}
+ \frac{1}{12} R \ln(\xi_i-\xi_j)^2 \Bigg\} \, ,
\end{multline}
where all curvature tensors are taken at $x_3$.
By item 1) of Theorem~2, we know that 
$C_I \sim_{\T,0} C_I^\T$. 

We next consider the first order perturbative 
contribution to a general OPE coefficient 
for a general triple product. By formula~\eqref{cijk} this is 
given by
\begin{multline}\label{luk}
C_{i_1 i_2 i_3}{}^{j}(x_1, x_2, x_3)_{(1)} =   
\i \int_M \Bigg[
\Psi^j(\O_{i_1}(x_1) \O_{i_2}(x_2) \R_1(\O_{i_3}(x_3), \L(y))) + 
{\rm cyclic}(1,2,3)\\
- \sum_{[k] \le \Delta} \Psi^j(\R_1(\O_k(x_3), \L(y)) 
\Psi^k(\O_{i_1}(x_1) \O_{i_2}(x_2) \O_{i_3}(x_3))
\Bigg] \chi(y) \, d\mu(y) \, .
\end{multline}
We are again interested in the case $\O_{i_1} = \O_{i_2} = \O_{i_3} = \O_j = \phi$. 
The constant $\Delta$ depends on the desired precision of the OPE
governed by $\delta$, and is given in Thm.~1. For our 
choice $\delta = 0$, we have we have to
choose $\Delta=3$. In this case, it can be seen that only 
$\O_k = \phi^3$ will make a contribution. 
Thus, the required retarded products and their
Wick expansion in eq.~\eqref{luk} are
\ben
\R_1(\phi^{n}(x), \phi^{m}(y)) = \sum_{k=0}^4
\frac{n!m!}{(n-k)!(m-k)!} \, r_k (x,y) \, :
\phi^{n-k}(x)\phi^{m-k}(y):_\h
\een
for $n=1,3$ and $m=4$. 
Consequently, we need to know $r_k$ for $k=1,3$. The method~\cite{hw2}
for defining the ``renormalized'' distribution gives 
$r_1 = \i \Delta_R$, where $\Delta_R$ is the retarded propagator, 
and
\bena
r_3 &=& -\frac{1}{32}v_0^3 \, A^2 \left(
\Theta(-t) 
  \frac{\ln (\sigma + \i 0 t)}{\sigma + \i 0 t}
\right)
\\
&+& \frac{3}{2} v_0^2 \sum_n v_{n+1} \frac{\sigma^n}{2^n n!} \, A\left(
\Theta(-t) \frac{\ln^2 (\sigma + \i 0 t)
+ \frac{1}{2} \ln (\sigma + \i 0 t)}{\sigma + i 0 t} 
\right)\nonumber\\
&+& \Theta(-t) \sum_{m,n} v_{m+1} v_{n+1} \frac{\sigma^{n+m}}{2^{n+m}n!m!}
\left( 3v_0 \frac{\ln^2(\sigma + \i 0
    t)}{\sigma + \i 0 t} + \sum_k v_{k+1} \frac{\sigma^k}{2^k k!} 
\ln^3 (\sigma+ \i 0 t)
\right) - \text{h.c.} \, , \nonumber 
\eena
where $t(x,y) = \tau(x) - \tau(y)$, for any time function $\tau:M \to
\mr$. The symbol $A$ stands for the operator\footnote{The usefulness of 
this operator lies in the identity
$A (\sigma^n) = 
4n(n+1)\sigma^{n-1}$.} 
\ben
A = \square + (\nabla^\mu \ln D)\nabla_\mu \, , 
\een
where $D$ is the VanVleck determinant defined above in eq.~\eqref{Ddef}.
The desired OPE coefficient can now be obtained. The result is 
\begin{multline}\label{CI}
C(x_1, x_2, x_3)_{(1)} = 
\i \kappa \int_M [ 
H(x_1, y) H(x_2, y) r_1(x_3, y) \\
+{\rm cyclic}(1,2,3)
- r_3(x_3, y) ] \chi(y) \, d\mu(y) \, , 
\end{multline}
where $H$ is the local Hadamard parametrix. To get the 
desired representer $C^\T_I$ provided by Theorem~2, 
we again use the definitions eqs.~\eqref{mellin}
and~\eqref{Ctdef} corresponding to our choice $\delta=0$.
Thus, we must compose $C_I$ with $\psi_\T(\epsilon)$, take a 
Mellin transform, and extract the residue at $0,-\i,-2\i$
in the complex $z$-plane.
To do the Mellin transform~\eqref{mellin}, we perform first a 
short distance expansion of the integrand
in~\eqref{CI} around $x_3$, using the corresponding expansions~\cite{deWitt} 
of the quantities $\sigma, v_n$ present in $H, r_i$.
This short distance expansion will lead to a sum of terms, 
each of which is a curvature 
polynomial at $x_3$, times a Minkowski distribution in $\xi_1, \xi_2$
and the Riemannian normal coordinates of $y$. 
We may also set $\chi = 1$, since the residue are
independent of the particular choice of $\chi$, as proven in Thm.~2. 
Thus, the computation of the Mellin-transform reduces to ordinary 
Minkowski integrals, times curvature polynomials at $x_3$. 
Only a finite number of these terms will give rise to poles 
at $0,-\i,-2\i$, and so all others can be discarded.
 
Let us consider in detail the pole $-2\i$. For this pole, 
the Minkowski space integrals that contribute
can be reduced to the integral given in~\cite{luperez} 
using some standard ``$\i 0$-identities'' for distributions:
\begin{multline}
C^\T(x_1, x_2, x_3)_{(1)} = \\
\frac{\kappa}{2^6 \pi^4} \frac{1}{\rho} \left[
{\rm Cl}_2 \left( 2\arctan \frac{\rho}{\delta_1} \right)
+
{\rm Cl}_2 \left( 2\arctan \frac{\rho}{\delta_2} \right)
+
{\rm Cl}_2 \left( 2\arctan \frac{\rho}{\delta_3} \right) + \dots
\right]\, , 
\end{multline}
where the dots stand for the contributions from the poles 
$0,-\i$. Here, we have set 
\ben
\rho = \sqrt{|\delta_1 \delta_2 + \delta_2 \delta_3 + \delta_3 \delta_1|}
\een
as well as
\ben
\delta_1 = (\xi_2-\xi_3)(\xi_3-\xi_1), \quad
\delta_2 = (\xi_1-\xi_2)(\xi_2-\xi_3), \quad
\delta_3 = (\xi_3-\xi_1)(\xi_1-\xi_2), 
\een
and as before it is assumed that the points $\xi_i \in \mr^4$ are
pairwise spacelike, and $\xi_3 = 0$. The function $Cl_2$ is the 
Clausen function~\cite{lewin}. 
The arguments of these functions in the above expression are given by twice 
the angles of a triangle with sides of 
length $\sqrt{(\xi_1-\xi_2)^2}, \sqrt{(\xi_2-\xi_3)^2}$ and 
$\sqrt{(\xi_3-\xi_1)^2}$, and $\rho$ represents the area of that
triangle, see the figure. 

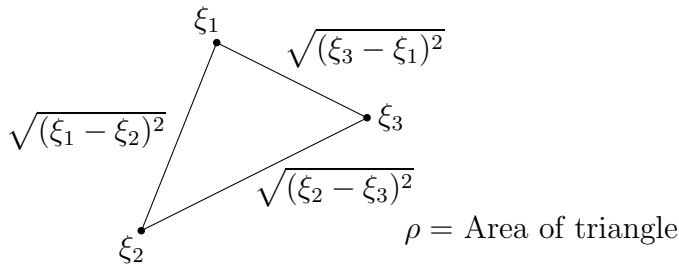
\setlength{\unitlength}{1cm}
\begin{center}
\begin{picture}(6,4)
\put(1,0.5){\line(2,1){3}}
\put(1,0.5){\circle*{0.1}}
\put(4,2){\line(-2,1){2}}
\put(4,2){\circle*{0.1}}
\put(2,3){\line(-2,-5){1}}
\put(2,3){\circle*{0.1}}
\put(0.7,0.15){$\xi_2$}
\put(4.15,1.9){$\xi_3$}
\put(1.7,3.2){$\xi_1$}
\put(2.9,2.8){$\sqrt{(\xi_3-\xi_1)^2}$}
\put(-0.8,1.7){$\sqrt{(\xi_1-\xi_2)^2}$}
\put(2.5,0.95){$\sqrt{(\xi_2-\xi_3)^2}$}
\put(4.5,0.4){$\text{$\rho=$ Area of triangle}$}
\end{picture}
\end{center}

Thus, the result for $C_I^\T$ manifestly has the simple form claimed in 
the scaling expansion in Theorem~2. There are no curvature terms in
the terms that we have displayed, but those arise from the other poles
$0,-\i$. These contributions may be obtained in closed form using the 
Minkowski integrals of~\cite{davy}.
The end result is a sum of terms 
$\sum_a R_{\mu\nu\sigma\rho} W^{\mu\nu\sigma\rho}_a(\xi_1, \xi_2, \xi_3)$, 
where $W^{\mu\nu\sigma\rho}_a$ are
Lorentz invariant distributions in the Riemannian 
normal coordinates $\xi_i$ of $x_i$ around $x_3$. 
However, the expressions for these distributions 
are rather complicated and will therefore be given 
elsewhere~\cite{holprep}. 

It should be clear that our method is not confined to the above 
example, but in principle only limited by the ability to perform 
complicated Minkowski integrals of Feynman type.

\paragraph{\bf Acknowledgements:} I have benefitted from 
conversations with D. Buchholz, N. Nikolov, and R.M. Wald.
I would like to thank M. D\"utsch for reading an earlier 
version of the manuscript.

\end{document}